\documentclass[times,twocolumn,final]{elsarticle}

\usepackage{medima}
\usepackage{framed,multirow}

\usepackage[algo2e,boxed,algoruled]{algorithm2e}
\usepackage{amssymb}
\usepackage{latexsym}
\usepackage{amsmath,amsfonts}
\usepackage{algorithmic}
\usepackage{graphicx}
\usepackage{nth}
\usepackage{textcomp}
\usepackage{stackengine}
\usepackage{subfigure}
\usepackage{framed,multirow}
\usepackage{booktabs}
\usepackage{tabularx}
\newcommand\setrow[1]{\gdef\rowmac{#1}#1\ignorespaces}
\usepackage{threeparttable}
\usepackage{dblfloatfix}
\DeclareMathOperator*{\argminB}{argmin}   
\usepackage{url}
\usepackage{xcolor}

\usepackage{hyperref}

\definecolor{newcolor}{rgb}{.8,.349,.1}

\begin{document}


\begin{frontmatter}

\title{Unsupervised diffeomorphic cardiac image registration using parameterization of the deformation field}%

\author[1,2]{Ameneh \snm{Sheikhjafari}\corref{cor1}}
\cortext[cor1]{Corresponding author: 
  Tel.: +1-412-721-3083;}
\ead{sheikhja@ualberta.ca}

\author[2]{Deepa \snm{Krishnaswamy}}
\ead{deepa@ualberta.ca}

\author[2]{Michelle \snm{Noga}}
\ead{mnoga@ualberta.ca}

\author[1]{Nilanjan \snm{Ray}\fnref{fn1}}
\ead{nray1@ualberta.ca}

\author[2]{Kumaradevan \snm{Punithakumar}\fnref{fn1}}
\ead{punithak@ualberta.ca}
\fntext[fn1]{Nilanjan Ray and Kumaradevan Punithakumar contributed equally as co–senior authors to this work.}
\address[1]{Department of Computing Science, University of Alberta, Edmonton, AB T6G 2G8, Canada}
\address[2]{Department of Radiology and Diagnostic Imaging, Edmonton, AB T6G 2G8, Canada}


\begin{abstract}
This study proposes an end-to-end unsupervised diffeomorphic deformable registration framework based on moving mesh parameterization. Using this parameterization, a deformation field can be modeled with its transformation Jacobian determinant and curl of end velocity field. The new model of the deformation field has three important advantages; firstly, it relaxes the need for an explicit regularization term and the corresponding weight in the cost function. The smoothness is implicitly embedded in the solution which results in a physically plausible deformation field. Secondly, it guarantees diffeomorphism through explicit constraints applied to the transformation Jacobian determinant to keep it positive. Finally, it is suitable for cardiac data processing, since the nature of this parameterization is to define the deformation field in terms of the radial and rotational components. The effectiveness of the algorithm is investigated by evaluating the proposed method on three different data sets including 2D and 3D cardiac MRI scans. The results demonstrate that the proposed framework outperforms existing learning-based and non-learning-based methods while generating diffeomorphic transformations.

\end{abstract}


\end{frontmatter}

\vspace*{200pt}
\textit{Preprint submitted to Medical Image Analysis}
\section{Introduction}
Deformable image registration plays a fundamental role in a variety of medical image analyses such as image guided-surgery \cite{ han2021fracture}, visual stabilization \cite{sheikhjafari2015robust}, reconstruction \cite{liu2021rethinking}, and the construction of many other image analysis problems \cite{krebs2019learning, haskins2019deep, sheikhjafari20153d}.
Many existing state-of-the-art deformable registration methods use traditional iterative algorithms,
such as standard symmetric normalization (SyN) \cite{wu2018automated} and log-domain based transformation \cite{mansi2011ilogdemons}. Due to the important properties such as folding-free and invertiblity \cite{dalca2018unsupervised} of diffeomorphic transformation, a wide range of researchers utilized diffeomorphisms by adding constraints to their formulation \cite{zhang2015finite, avants2008symmetric, vercauteren2008symmetric, punithakumar2013regional}.
These traditional algorithms are computationally expensive and do not learn the features from data to be registered. Recently, \cite{sheikhjafari2022training} proposed a convolutional neural network (CNN) to model the optimization problem for deformable registration and shared the parameters through a temporal sequence. However, they still establish the displacement field via iterative optimization between images. Even though traditional deformable image registration techniques can generate promising mappings between images, most of these methods require users to identify parameters that match the characteristics of the problem and manually adjust regularization terms for each application to obtain accurate results.  

In recent years, the popularity of learning-based registration algorithms has been increasing due to the lower computational costs and execution times \cite{krebs2018unsupervised}. In supervised-learning methods, a CNN is trained using examples of medical images along with their ground truth transformations to predict the transformations directly on test images \cite{rohe2017svf, cao2017deformable}. Even though the accuracy of these approaches is considerable, their performance is highly dependent on the quality of the ground truth \cite{sang2020imposing}. One of the most significant challenges in applying the supervised methods to medical imaging applications is that the actual ground truth of a desired neural network output is not often available. 

With that limitation in mind, several unsupervised learning-based image registrations have been proposed. Most of the unsupervised approaches use spatial transformer layers (STN) to warp the moving image in a differentiable way. In this way, an optimization can be performed by using a similarity metric based on the warped image \cite{jaderberg2015spatial, de2017end, de2019deep, balakrishnan2018unsupervised, sheikhjafari2018unsupervised}.

When image registration is stated as an optimization of a similarity metric alone, it is commonly understood as an ill-posed problem. To tackle this problem, a regularization approach is commonly used. Without regularization, this may result in multiple and physically non-plausible solutions. For instance, it might lead to tissue folding and tearing of anatomical structures in images. 
Aside from that, while unsupervised approaches can perform well in minimizing a similarity metric between warped moving images and fixed images, important properties such as symmetry, diffeomorphism, and regularity of the retrieved deformation fields are still unclear and missing.   \cite{rohlfing2011image,haber2004numerical}.

Inspired by \cite{punithakumar2015right,punithakumar2017gpu, chen2010parameterization}, we tackle aforementioned issues with the help of moving mesh parameterization which was originally designed to generate a suitable grid for solving partial differential equations  \cite{haber2004numerical}. We propose a ConvNet method based on unsupervised learning for deformable cardiac registration, which formulates the deformation field by the moving mesh approach. This parameterization naturally leads to a formulation of diffeomorphic image registration as a constrained optimization problem. It also bypasses the need for an explicit regularization term and the corresponding weight in the cost function. Such a strategy has been adopted in the demons algorithm, where unconstrained optimization is followed by Gaussian filtering to impose a smoothness constraint. Using the moving mesh grid generation, we can define a deformation field with its transformation Jacobian determinant and curl of end velocity field which make it appealing to image registration \cite{punithakumar2015right,punithakumar2017gpu}. The new formulation of the deformation field ensures diffeomorphic properties on the deformation field by explicitly applying constraints on transformation Jacobian determinant to keep it positive. 
Since the heart motion could be decomposed of radial expansion and twisting \cite{garreau2006assessment}, defining the deformation field in terms of radial and rotational components makes this formulation suitable for cardiac analysis \cite{bijnens2012myocardial}.

\section{Methodology}
Most of the learning-based algorithms formulate the deformable registration problem as the minimization of the following equation:
\begin{equation}
     \phi^{\ast}=
     \argminB_{\phi} {L(I_F,I_M \circ \phi(\xi))}
     \label{eq6:0}
\end{equation}
where $\xi$ denotes the pixel location in the image domain $\Omega$, $\phi: \Omega \,\to\, \Omega$ denotes the transformation function, and the dissimilarity metric is denoted by $L(.)$.
With the above formulation, introducing a regularization is necessarily to obtain a unique solution. Without regularization, this may result in multiple physically non-plausible solutions.

In our setting, we tackle these issues with the help of the moving mesh parameterization.
\subsection{Moving Mesh Grid Generation}
To avoid adding extra terms to the above formulation and having a unique solution, more constraints are required to be added using a monitor function $\mu$ and curl of end velocity field $\gamma$. 

First a continuous monitor function is defined and constrained by: 

\begin{equation} \label{eq6:1}
\int_\Omega \mu = |\Omega|. 
\end{equation}

The goal here is to find a transformation $\phi_1$: $\Omega \,\to\, \Omega$, ${\partial \Omega} \,\to\, {\partial \Omega}$ such that the transformation Jacobian determinant $J_{\phi_1}(\xi)$ is equal to the monitor function $\mu$ : 

\begin{equation} \label{eq6:2}
J_{\phi}(\xi) = det \nabla \phi_1(\xi) = \mu(\xi).
\end{equation} 

To find the transformation $\phi_1$ which satisfies \ref{eq6:2}, the following steps need to be taken,

\textit{Step 1}: A vector field $V(\xi)$ is defined such that:
\begin{equation}\label{eq6:3}
   {div} \: {V(\xi)} =  \mu(\xi) - 1.
\end{equation}

\textit{Step 2}: A velocity vector field based on artificial-time is then constructed from $V(\xi)$: 
\begin{equation}\label{eq6:4}
    V_t(\xi) = \frac{V(\xi)}{t + (1-t)\mu(\xi)}, t \in [0,1]
\end{equation}
The desire transformation $\phi_1$ can be found by solving the following ordinary differential equation (ODE) at $t=1$, $\phi_1(\xi) = \psi(\xi,t=1)$ where $\psi(\xi,$t=0$)=\phi_0(\xi)$
\begin{equation}\label{eq6:5}
    \frac{\psi(\xi,t)}{dt} = V_t(\psi(\xi,t)), t \in [0,1],
\end{equation}
Where $\phi_0(\xi)$ is the identity mapping and $det \nabla \phi_0(\xi) = 1$ and $\phi_0(\xi) = \xi$. Since the $\phi_1(\xi)$ is the desire transformation that we are looking for, we drop the subscript and use $\phi(\xi)$ for the rest of the paper.
\begin{algorithm2e*}[hpt]
\caption{Moving Mesh based deformable registration }
\label{alg6:cnn_alg}
\SetAlgoLined
\KwIn{Given two 2D/3D pair of images, fixed image $I_F$ and moving image $I_M$. The upper bound $\tau_{ub}$ and lower bound $\tau_{lb}$ of the transformation Jacobian determinant}
\KwOut{Deformation field $\phi$}
\DontPrintSemicolon
Step 1: Pass the input to the CNN to compute $\mu(\xi)$ and $V(\xi)$;

Step 2: Impose constraints from \eqref{eq6:7} for each pixel location $\xi \in \Omega:$ \;


$\mu(\xi) \leftarrow \dfrac{|\Omega|}{\sum_{\xi\subset\Omega}^{}\mu(\xi)}$\;

Step 3: Compute a curl of velocity field $V(\xi)$ that satisfies \eqref{eq6:3} and compute the deformation field $\phi$\;
Step 4: Compute the loss function\;
Step 5: Update the $\mu$ and $V(\xi)$ using back-propagation\; 

\end{algorithm2e*}
The main problem is how to find $V(\xi)$ such that $div V(\xi) = \mu(\xi) - 1$. There are different methods to solve this problem such as the div-curl system. To solve the problem with the div-curl system, we need to find the divergence and curl at each point and set up the div-curl system of equations for each point. By solving this system we can reconstruct a differentiable and invertible transformation.
\begin{equation} \label{eq6:6}
\begin{cases}
\begin{aligned}
div V(\xi)  &= \mu(\xi) - 1\\ 
curl V(\xi) &= \gamma(\xi).  \\
\end{aligned}
\end{cases}
\end{equation}

To have a unique $\phi$ a constraint need to be applied to the div of the vector field $V(\xi)$, \ref{eq6:6}.

The generated transformation $\phi$ now can be parameterized with transformation Jacobian determinant and the curl of the end velocity field.
\subsection{Diffeomorphic Image Registration}
Using the above parameterization, the diffeomorphic image registration can be formulated as a constrained optimization problem. 

Let $I_F$ and $I_M$ be 2D/3D fixed and moving images/volumes, defined over $\Omega \,\to\, \mathbb{R}^2$/$\Omega \,\to\, \mathbb{R}^3$. We need to find $\mu(\xi)$ and $\gamma(\xi)$ $\forall \xi \in \Omega$, that optimize a similarity metric $L_{Sim}$ between the warped moving image and fixed image, subject to the following constraints:

\begin{equation} \label{eq6:7}
\begin{cases}
\begin{aligned}
\int \mu(\xi){d\xi} = |\Omega| \\
\tau_{ub} > \mu(\xi) > \tau_{lb}
\end{aligned}
\end{cases}
\end{equation}

where the $\tau_{ub}$ is the upper bound and $\tau_{lb}$ is the lower bound  of the transformation Jacobian determinant which were set by the user. The $\tau_{lb} > 0$ guarantees the diffeomorphism.

\begin{figure*}[htbp]
	\centering
	\includegraphics[width=1\textwidth]{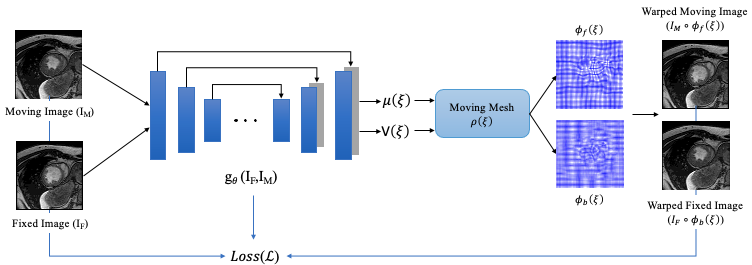}
	\caption{Overview of end-to-end unsupervised architecture. The ConvNet $g_\theta(I_F,I_M)$ takes the input fixed image($I_F$) and moving image($I_M$) and outputs the transformation Jacobian determinant $J_{\phi}(\xi) = \mu(\xi)$, and the vector field $V(\xi)$. Then the diffeomorphic forward and backward transformations $\phi_f$ and $\phi_b$ are computed using the moving mesh approach. Finally, the moving and fixed images are warped using $\phi_f$ and $\phi_b$.
}
	\label{fig6:net}
\end{figure*} 
\subsection{Numerical Methods}

\subsubsection{2D Div-curl solver}
We represent the deformation field by divergence and curl (div-curl) system representation \cite{Cheng1989} \eqref{eq6:6}. To find $V(\xi)$ under the null condition we converted the \eqref{eq6:6} into a set of Poisson equations as follows and used a Fast Fourier Transform (FFT) based Poisson solver. As shown in \eqref{eq6:9_1} the radial component is given by $F^1$ and the rotational components is given by $F^2$:
\begin{equation} \label{eq6:9_1}
\begin{cases}
\Delta V_x &= \mu_x - \gamma_y = F^1, \\
\Delta V_y &= \mu_y + \gamma_x = F^2, \\

\end{cases}
\end{equation}

\subsubsection{3D Div-curl solver}
The div-curl system for the 3D case is given in Equation \eqref{eq6:6}. Where the divergence of the deformation field represents the radial motion while the curl operator represents the rotation of the media around every point. The 3D operator directly extends from the 2D curl, where each rotational component represents the rotational motion of the deformation field about each of the three axes. As it shown in \eqref{eq6:10} the radial component is given by $f^1$ and the three rotational components are given by $f^2$, $f^3$ and $f^4$. For the 3D version, there ate three unknowns ($V_x, V_y, V_z$) with four scalar equations which makes this system overdetermined. Furthermore, a dummy variable $\theta$ is introduced to solve the system. (please check \cite{liu2006new} for more details.)
\begin{equation} \label{eq6:10}
\begin{cases}
\begin{aligned}
div V &=  \displaystyle\frac{\partial V_x}{\partial x} + \frac{\partial V_y}{\partial y} + \frac{\partial V_z}{\partial z} = f^1 \\ 
curl_x V &= \displaystyle\frac{\partial \theta}{\partial x}+\frac{\partial V_z}{\partial y} - \frac{\partial V_y}{\partial z} = f^2 \\
curl_y V &= \displaystyle\frac{\partial \theta}{\partial y}+\frac{\partial V_x}{\partial z} - \frac{\partial  V_z}{\partial x} = f^3 \\
curl_z V &= \displaystyle\frac{\partial \theta}{\partial z}+\frac{\partial V_y}{\partial x} - \frac{\partial  V_x}{\partial y} = f^4. 
\end{aligned}
\end{cases}
\end{equation}
Similar to the 2D version, we converted the \eqref{eq6:10} into a set of Poisson equations as follows:
\begin{equation} \label{eq6:11}
\begin{cases}
\Delta V_x &= f^1_x + f^3_z - f^4_y = F^1, \\
\Delta V_y &= f^1_y + f^4_x - f^2_z = F^2, \\
\Delta V_z &= f^1_z + f^2_y - f^3_x = F^3. \\
\end{cases}
\end{equation}

Then the Euler method with arbitrary time steps is used to compute the transformation $\phi$ from $V(\xi)$ via \eqref{eq6:4} and \eqref{eq6:5}. For derivation and numerical implementation details, we refer the reader to \cite{liu2006new}

\subsection{Data driven parameter computation}
Despite the traditional methods that iteratively and manually compute the parameters and update the gradient \cite{chen2010parameterization,punithakumar2017gpu} which are time-consuming, we use an unsupervised CNN and back-propagation Algorithm \ref{alg6:cnn_alg}. 
In the proposed framework, the network parameters are learnt in an unsupervised fashion and a diffeomorphic deformation field is generated by moving mesh parameterization Figure \ref{fig6:net}. 

As shown in Figure \ref{fig6:net}, the network takes $I_F$ and $I_M$ as input and outputs the monitor function $\mu(\xi)$ and the velocity vector filed $V(\xi)$. Then using the curl of end velocity and a div-cur system a diffeomorphic transformation $\phi$ is computed. To establish the uniqueness of the solution the Dirichlet boundary condition is used \cite{zhou2006uniqueness}. Additionally, a diffeomorphism, which is corresponded to a positive transformation Jacobian determinant, is enforced explicitly via the monitor function \cite{liu2006new}. All of the steps are designed to be differentiable and the network parameters are learnt using stochastic gradient descent optimization. 

\subsection{Registration}
To train the framework a set of pair images $(I_F, I_M)$ were given. Then using the monitor function and curl of end velocity, the desire $\phi$ was computed. Finally, the moving image was warped to have the minimum dissimilarity with fixed image $I_F$. For each pair of image, we simultaneously calculated the forward transformation which registers the fixed image $I_F$ to moving image $I_M$ and the backward transformation which registers the moving image $I_M$ to fixed image $I_F$.
A symmetric loss function is used as follows:
\begin{equation}
     \phi^{\ast}=
     \argminB_{\theta,\mu , \gamma} \{ w \times L(I_F,I_M \circ \phi_f ) + w \times L(I_M,I_F \circ \phi_b) \}
     \label{eq6:eq10}
\end{equation}
Where $\phi_f$ is the forward transformation and $\phi_b$ is the backward transformation.

The registration process is performed pairwise on both 2D images and 3D volumes. In the cardiac data sets, the end-diastolic and end-systolic images are passed to the proposed framework as input to compute the forward transformation $\phi_f$ and the reverse transformation $\phi_b$.
For the 2D version the mean squared error (MSE) and for the 3D version the normalise cross correlation (NCC) is used as dissimilarity metric.

\section{Experiments}
We perform a series of experiments to evaluate the registration accuracy of the proposed diffeomorphic CNN method against the state-of-the-art methods. The evaluations were performed over three data sets consisting of clinical 2D cardiac MR images to assess the performance of the 2D version of our method. We also evaluated the 3D version of the proposed framework using ACDC data set in 3D.
\subsection{Data sets}
The following three data sets are considered in this study:
\paragraph{
Automated Cardiac Diagnosis Challenge (ACDC) \cite{bernard2018deep}}  
This data set contains multiple temporal 2D short-axis cardiac cine MRI sequences acquired from 100 patients and is one of the publicly available data sets for cardiac MRI assessment. 
The spatial resolution varies from $1.37$ to $1.68$ $mm^2 / pixel$ with a slice thickness of 5 mm to 8 mm (in average 5mm). The testing set contained 20 cases of each of the following cardiac diseases: dilated cardiomyopathy (DCM), hypertrophic cardiomyopathy (HCM), previous myocardial infarction (MINF), abnormal right ventricle (RV) and healthy (Normal). The images are cropped to a size of $128 \times 128$, and padded the third dimension to $16$ for the 3D voxels.

\paragraph{The Sunnybrook Cardiac Challenge data (SCD) \cite{radau2009evaluation}} 
This data set contains multiple temporal 2D short-axis cardiac cine MRI scans acquired from 45 patients. Each cine sequence includes 20 frames to cover the cardiac cycle. The data set is equally divided into 15 patient scans for training, 15 patient scans for validation, and 15 patient scans for testing. The image resolution is $256\times256$, with a pixel spacing of 1.25 mm and slice thickness of 8 mm. 
\paragraph{Left Atrium (LA)} This data set includes 100 temporal 2D long-axis cine MRI steady-state sequences from the 2, 3 and 4-chamber views, acquired from the University Alberta Hospital. Each cycle includes 25 or 30 frames with image resolutions $176\times189$ -- $256\times208$ and image spacing $1.445 - 1.795$ mm. The ground truth manual segmentation is initially performed by a medical student and edited by an experienced radiologist. The 2ch, 3ch and 4ch are used in the rest of the paper to denote 2, 3 and 4-chamber sequences, respectively.
The results are compared on end-diastolic and end-systolic frames.
\subsection{Quantitative Evaluation Metrics}
The proposed method is evaluated quantitatively using four metrics, namely, Dice metric (DM), Hausdorff distance (HD in mm), determinant of Jacobian of the deformation field $\det(J)$, and reliability $R(d)$.

\textbf{Dice Metric} 
The DM \cite{dice1945measures} is a segmentation-based metric to measure the similarity (overlap) between two regions, warped moving and fixed images. Where the Dice score of 1 indicates complete overlap and Dice score of 0 indicates no overlap. The DM of two regions A and B is formulated as:
\begin{equation}
    \label{eq:Dice}
        DM(A,B) = \frac{2 |A \cap B|}{A+B}
\end{equation}%

\textbf{Hausdorff Distance}
The HD \cite{huttenlocher1993comparing} is another metric which measures the maximum deviation between two regions' contours. The HD between two contours $(C_A)$ and $C_B$ is formulated as:
\begin{equation}
    \label{eq:HD}
    \begin{split}
        \textnormal{HD}(C_A,C_B)= & \max(\max_{i}(\min_{j}(d(p^{i}_{A},p^{j}_{B}))),\\
        & \max_{j}(\min_{i}(d(p^{i}_{A},p^{j}_{B}))))
    \end{split}
\end{equation}%
where $p^{i}_{A}, p^{j}_{B}$ denote the set of all the points in $C_A$ and $C_B$ respectively. The term $d(\cdot)$ denotes the Euclidean distance.

\textbf{Reliability:} We also evaluated the performance of the proposed algorithm using a reliability function computed based on DMs for each data set. The complementary cumulative distribution function is defined for each $d \in [0, 1]$ as the probability of obtaining $DM$ higher than $d$ overall volumes.
\begin{equation}
    \label{eq:relability}
    \begin{split}
        R(d) &= P_r(Dice > d)\\
        & = \frac{\textnormal{\# Images segmented with DM higher than d}}{\textnormal{total number of images}}.
    \end{split}
\end{equation}
$R(d)$ measures how reliable the algorithm is in yielding accuracy $d$. 

\textbf{det(J):} To analyze deformation regularity in different algorithms, we calculate the determinant of the Jacobian $\det(J)$ \cite{ashburner1999high}. If the value of $\det(J)$ equals $1$, the area remains constant after the transformation, whereas the value smaller or larger than $1$ indicates the local area shrinkage or expansion, respectively. The negative value of $\det(J)$ implies that local folding and twisting have occurred, which are physically not realizable and mathematically not invertible \cite{dalca2018unsupervised}.
\begin{table*}[htbp]
\centering
\caption{Quantitative evaluation of the results for cardiac MRI registration on the 2D ACDC data set. The following metrics are reported for each method: The Dice score $Dice$ (mean$\pm$ standard deviation), Hausdorff distance $HD$, the percentage of the number of pixels with negative Jacobian determinant $\% |J_\theta|<0$, and reliability $R(0.75)$. Smaller values of $HD$ and larger values of $Dice$ indicate more accurate results. Also the smaller $\% |J_\theta|<0$ indicates less mesh folding. The higher probability values of $R(0.75)$ show that more patients have the dice score higher or equal to $\%0.75$. Values that are highlighted in bold indicate the metric that gave the best performance compared to the other algorithms.}
\label{ch6:ACDCtable_2dreg}
\setlength{\tabcolsep}{5pt}
\begin{tabular}{p{180pt} p{90pt} p{70pt} p{70pt}p{40pt}}
\hline
Method&
Dice& 
HD&
$\% |J_\theta|<0$ &
$R(0.75)$\\
\hline
Undeformed & 0.71 $\pm$ 0.15 & 10.1 & --& --  \\
Demon\cite{yoo2002engineering}& 0.76 $\pm$ 0.10 & 8.3 & 0.27 & 0.36\\
SyN\cite{avants2008symmetric} & 0.80 $\pm$ 0.09 &8.1& 0.28& 0.66\\
LPM\cite{krebs2019learning}& 0.79 $\pm$ 0.10 & 7.6  & 0.38 &0.46\\
MM\cite{punithakumar2017gpu} & 0.83 $\pm$ 0.15 & 5.64& 0 & 0.81\\
Elastix\cite{marstal2016simpleelastix} & 0.84 $\pm$ 0.14 &  4.51& 0.12 &  0.82\\
\setrow{\bfseries}Proposed Method &\setrow{\bfseries}0.88 $\pm$ 0.11 &\setrow{\bfseries} 3.85& \setrow{\bfseries}0& \setrow{\bfseries}0.89\\
\hline
\end{tabular}
\end{table*}
\begin{table*}[htbp]
      \caption{
      Quantitative evaluation of the results for cardiac MRI registration on the 2D LA data set. The following metrics are reported for each method: The Dice score $Dice$ (mean$\pm$ standard deviation), Hausdorff distance $HD$, the percentage of the number of pixels with negative Jacobian determinant $\% |J_\theta|<0$, and reliability $R(0.75)$. The 2ch, 3ch and 4ch stand for the 2, 3 and 4-chamber. Values in bold indicate the best performance.}
    \begin{minipage}{1\linewidth}
    \vspace{3pt}
    	\centerline{(a) 2ch}\medskip
        \centering
        \begin{tabular}{p{180pt} p{90pt} p{70pt} p{70pt}p{40pt}}
        \hline
        \small{Methods} &\small{ Dice} & \small{HD }& \small{$\% |J_\theta|<0$} & \small{$R(0.75)$ }\\
        \hline
        Undeformed  &  0.79 $\pm$ 0.07 & 7.37 & -- & --\\
        Demons\cite{mccormick2014itk} &  0.84 $\pm$ 0.08 & 7.41 & 0.38 & 0.89\\
        SyN\cite{avants2008symmetric} &  0.87 $\pm$ 0.06&\setrow{\bfseries} 6.38& 0.18 & 0.95\\
        MM\cite{punithakumar2017gpu} &  0.84 $\pm$ 0.06 & 6.58 & 0  & 0.92\\
        Elastix\cite{marstal2016simpleelastix} &  0.82 $\pm$ 0.11& 7.28 & 0.28 & 0.74\\
        
        \setrow{\bfseries}Proposed Method &\setrow{\bfseries} 0.88 $\pm$ 0.04 &  6.54 & \setrow{\bfseries} 0&\setrow{\bfseries} 0.95\\
        \hline
        \end{tabular}
    \end{minipage}%

    \begin{minipage}{1\linewidth}
    \vspace{3pt}
    \centerline{(b) 3ch}\medskip
        \centering
        \begin{tabular}{p{180pt} p{90pt} p{70pt} p{70pt}p{40pt}}
        \hline
        \small{Methods} &\small{ Dice} & \small{HD }& \small{$\% |J_\theta|<0$} & \small{$R(0.75)$ }\\
        \hline
        Undeformed  & 0.78 $\pm$ 0.08 & 7.70 & --&--\\
        Demons\cite{mccormick2014itk} &  0.85 $\pm$ 0.06 & 7.33 &0.36&0.94\\
        SyN\cite{avants2008symmetric} &  0.86 $\pm$ 0.13 & 7.52 & 0.21&0.93\\
        MM\cite{punithakumar2017gpu} & 0.83 $\pm$ 0.06& 6.48 & 0&0.88\\
        Elastix\cite{marstal2016simpleelastix} &  0.86 $\pm$ 0.10& 6.82 &0.26&0.9\\
        
        \setrow{\bfseries}Proposed Method & \setrow{\bfseries} 0.87 $\pm$ 0.05& \setrow{\bfseries}6.3 &\setrow{\bfseries}0&\setrow{\bfseries}0.94\\
        \hline
        \end{tabular}
    \end{minipage}%
    
    \begin{minipage}{1\linewidth}
    \vspace{3pt}
    \centerline{(c) 4ch}\medskip
        \centering
        \begin{tabular}{p{180pt} p{90pt} p{70pt} p{70pt}p{40pt}}
        \hline
        \small{Methods} &\small{ Dice} & \small{HD }& \small{$\% |J_\theta|<0$} & \small{$R(0.75)$ }\\
        \hline
        Undeformed  & 0.78 $\pm$ 0.09 & 8.66 &--&--\\
        Demons\cite{mccormick2014itk} &  0.82 $\pm$ 0.10 & 7.84&0.43&0.77\\
        SyN\cite{avants2008symmetric} & 0.84 $\pm$ 0.11 &7.51&0.20&0.86\\
        
        MM\cite{punithakumar2017gpu} & 0.83 $\pm$ 0.08 & 6.77& 0 & 0.87\\
        Elastix\cite{marstal2016simpleelastix} &  0.82 $\pm$ 0.10 & 7.56&0.38&0.64\\
        \setrow{\bfseries}Proposed Method &\setrow{\bfseries} 0.87 $\pm$ 0.05 &\setrow{\bfseries} 6.1&\setrow{\bfseries}0&\setrow{\bfseries}0.99\\
        \hline
    \end{tabular}
    \end{minipage}%
\label{ch6:lefttable_2dreg}
\end{table*}%
\begin{table*}[htbp]
\centering
\caption{Quantitative evaluation of the results for cardiac MRI registration on the 2D SCD data set. The following metrics are reported for each method: The Dice score $Dice$ (mean$\pm$ standard deviation), Hausdorff distance $HD$, the percentage of the number of pixels with negative Jacobian determinant $\% |J_\theta|<0$, and reliability $R(0.75)$. Smaller values of $HD$ and larger values of $Dice$ indicate more accurate results. Also the smaller $\% |J_\theta|<0$ indicates less mesh folding. The higher probability values of $R(0.75)$ show that more patients have the dice score higher or equal to $\%0.75$. Values that are highlighted in bold indicate the metric that gave the best performance compared to the other algorithms.}
\label{ch6:Sunnytable_2dreg}
\setlength{\tabcolsep}{5pt}
\begin{tabular}{p{180pt} p{90pt} p{70pt} p{70pt}p{40pt}}
\hline
Method&
Dice& 
HD&
$\% |J_\theta|<0$ &
$R(0.75)$\\
\hline
Undeformed & 0.62 $\pm$ 0.15 &16.02 & -- & --\\
Demons\cite{mccormick2014itk}  &  0.68 $\pm$ 0.18 &12.46 &0.4&0.36\\
SyN\cite{avants2008symmetric} & 0.81 $\pm$ 0.16 &  8.9 & 0.02 &0.70\\
LPM\cite{krebs2019learning}& 0.78 $\pm$ 0.08 & 7.6 &  0.38 &0.63\\
MM\cite{punithakumar2017gpu} & 0.72 $\pm$ 0.12 &  12.53 &  0 & 0.59 \\
Elastix\cite{marstal2016simpleelastix} &  0.79 $\pm$ 0.08 & 11.12 & 0.37& 0.62\\
\setrow{\bfseries}Proposed Method &\setrow{\bfseries}0.88 $\pm$ 0.09 & \setrow{\bfseries}5.25&\setrow{\bfseries}0 &\setrow{\bfseries} 0.90\\
\hline
\end{tabular}
\label{tab1}
\end{table*}
\begin{figure*}[htbp]
    
     \centering
    \begin{minipage}{0.99\linewidth}
		\centering
		\centerline{(a) 2ch}\medskip
	\end{minipage}
    \begin{center}
    \subfigure{\includegraphics[width=0.13\textwidth]{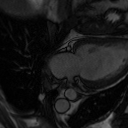}}
    \hspace{3pt}
    \subfigure{\includegraphics[width=0.13\textwidth]{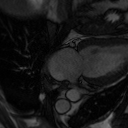}}
    \hspace{3pt}
    \subfigure{\includegraphics[width=0.13\textwidth]{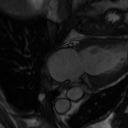}}
    \hspace{3pt}
    \subfigure{\includegraphics[width=0.13\textwidth]{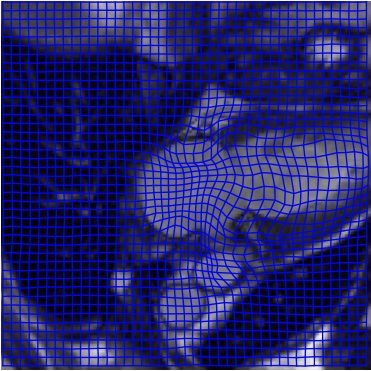}}
    \hspace{3pt}
    \subfigure{\includegraphics[width=0.13\textwidth]{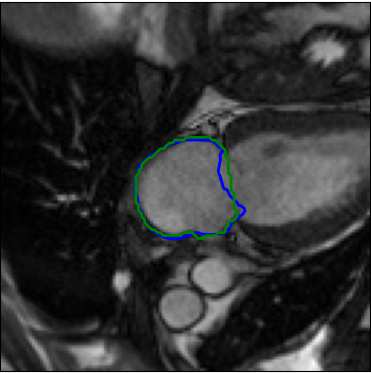}}
    
     \stackunder[5pt]{\includegraphics[width=0.13\textwidth]{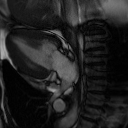}}{ED}
     \hspace{3pt}
     \stackunder[5pt]{\includegraphics[width=0.13\textwidth]{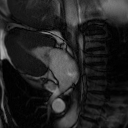}}{ES}
     \hspace{3pt}
     \stackunder[5pt]{\includegraphics[width=0.13\textwidth]{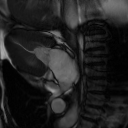}}{Warped ES}
     \hspace{3pt}
     \stackunder[5pt]{\includegraphics[width=0.13\textwidth]{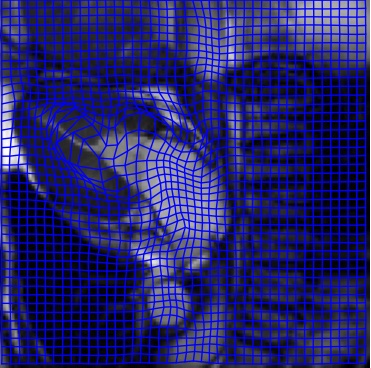}}{DF}
     \hspace{3pt}
     \stackunder[5pt]{\includegraphics[width=0.13\textwidth]{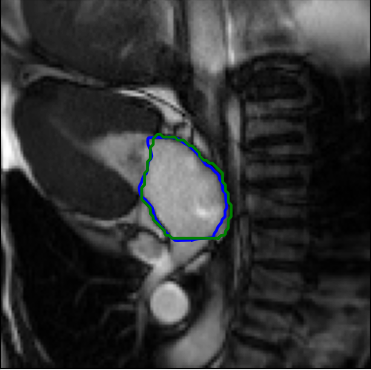}}{GT}
     \end{center}
 
    \begin{minipage}{0.99\linewidth}
		\centering
		\centerline{(b) 3ch}\medskip
	\end{minipage}
	     
    \begin{center}
    \subfigure{\includegraphics[width=0.13\textwidth]{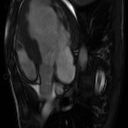}}
    \hspace{3pt}
    \subfigure{\includegraphics[width=0.13\textwidth]{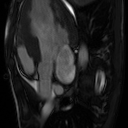}}
    \hspace{3pt}
    \subfigure{\includegraphics[width=0.13\textwidth]{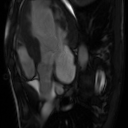}}
    \hspace{3pt}
    \subfigure{\includegraphics[width=0.13\textwidth]{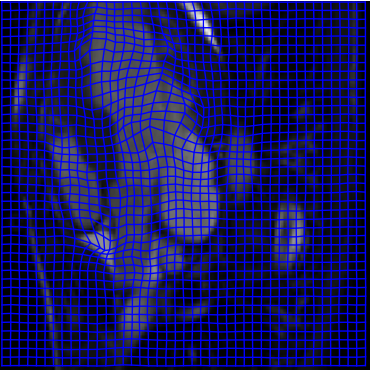}}
    \hspace{3pt}
    \subfigure{\includegraphics[width=0.13\textwidth]{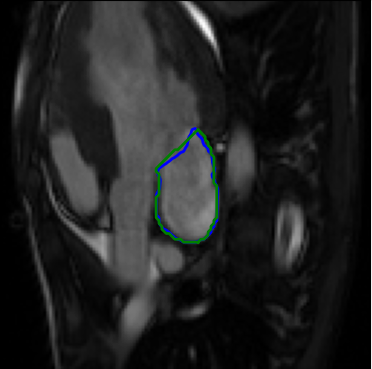}}
    
    \stackunder[5pt]{\includegraphics[width=0.13\textwidth]{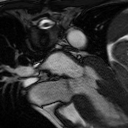}}{ES}
    \hspace{3pt}
    \stackunder[5pt]{\includegraphics[width=0.13\textwidth]{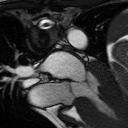}}{ED}
    \hspace{3pt}
    \stackunder[5pt]{\includegraphics[width=0.13\textwidth]{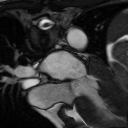}}{Warped ES}
    \hspace{3pt}
    \stackunder[5pt]{\includegraphics[width=0.13\textwidth]{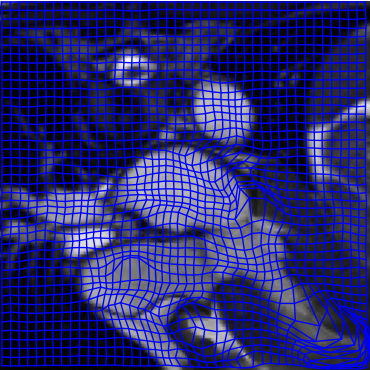}}{DF}
    \hspace{3pt}
    \stackunder[5pt]{\includegraphics[width=0.13\textwidth]{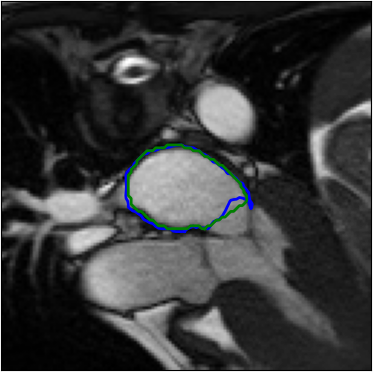}}{GT}
    \end{center}
    \begin{minipage}{0.99\linewidth}
		\centering
		\centerline{(b) 4ch}\medskip
    \end{minipage}
    \begin{center}
    \subfigure{\includegraphics[width=0.13\textwidth]{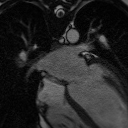}}
    \hspace{3pt}
    \subfigure{\includegraphics[width=0.13\textwidth]{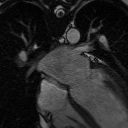}}
    \hspace{3pt}
    \subfigure{\includegraphics[width=0.13\textwidth]{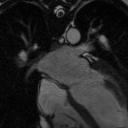}}
    \hspace{3pt}
    \subfigure{\includegraphics[width=0.13\textwidth]{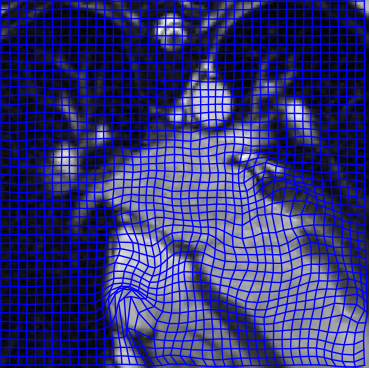}}
    \hspace{3pt}
    \subfigure{\includegraphics[width=0.13\textwidth]{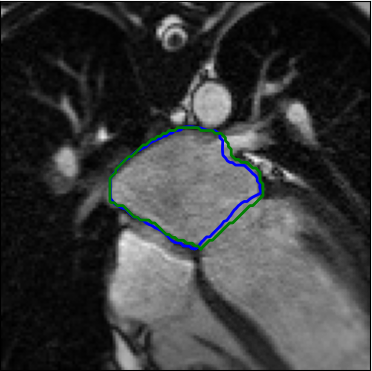}}
    
    \stackunder[5pt]{ \includegraphics[width=0.13\textwidth]{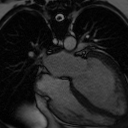}}{ES}
    \hspace{3pt}
    \stackunder[5pt]{\includegraphics[width=0.13\textwidth]{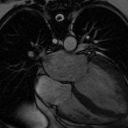}}{ED}
     \hspace{3pt}
      \stackunder[5pt]{\includegraphics[width=0.13\textwidth]{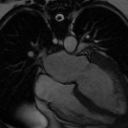}}{Warped ES}
      \hspace{3pt}
      \stackunder[5pt]{\includegraphics[width=0.13\textwidth]{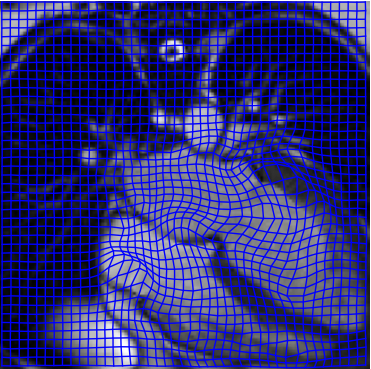}}{DF}
      \hspace{3pt}
     \stackunder[5pt]{\includegraphics[width=0.13\textwidth]{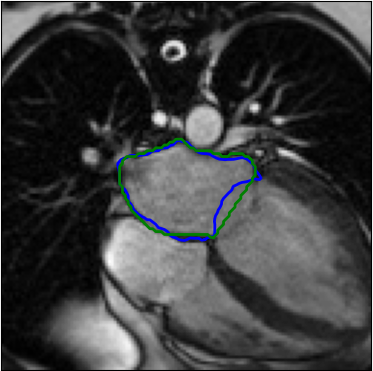}}{GT}
    \end{center}

      \begin{minipage}{0.7\linewidth}
		\centering
		\centerline{LA}\medskip
	\end{minipage}
     \caption{Samples of registered images on the left atrium data set with the corresponding deformation field grid (DF). The end-systolic (ES) frame is the moving image and end-diastolic (ED) frame is the fixed image. The warped ES of each row is shown in the third column. The last column labeled ground truth (GT) displays the true segmentation and the predicted segmentation, which are shown by the green line and blue line respectively. The 2ch, 3ch and 4ch stand for the 2, 3 and 4-chamber. }
     \label{fig6:la_2d_reg}
\end{figure*}

\begin{figure*}[htbp]
     \centering
     \subfigure{\includegraphics[width=0.13\textwidth]{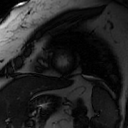}}
    \hspace{3pt}
    \subfigure{\includegraphics[width=0.13\textwidth]{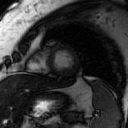}}
    \hspace{3pt}
    \subfigure{\includegraphics[width=0.13\textwidth]{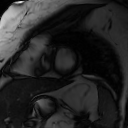}}
    \hspace{3pt}
    \subfigure{\includegraphics[width=0.13\textwidth]{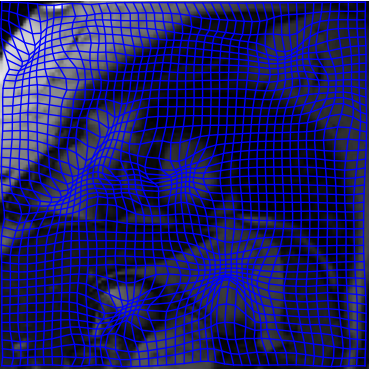}}
    \hspace{3pt}
    \subfigure{ \includegraphics[width=0.13\textwidth]{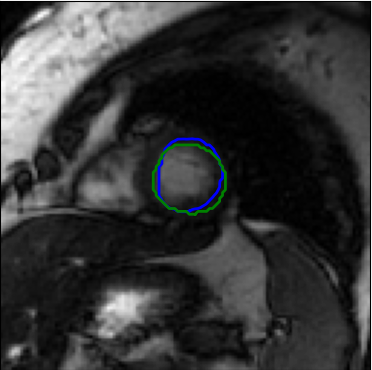}}
    \hspace{3pt}

    \subfigure{\includegraphics[width=0.13\textwidth]{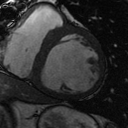}}
    \hspace{3pt}
    \subfigure{\includegraphics[width=0.13\textwidth]{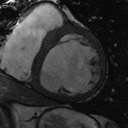}}
    \hspace{3pt}
    \subfigure{ \includegraphics[width=0.13\textwidth]{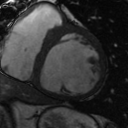}}
    \hspace{3pt}
    \subfigure{\includegraphics[width=0.13\textwidth]{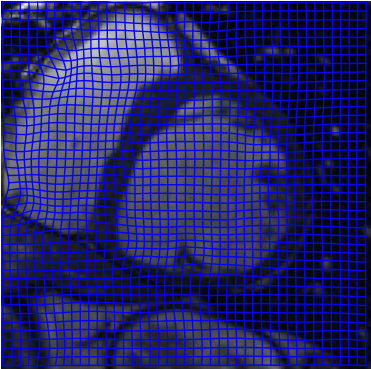}}
    \hspace{3pt}
    \subfigure{\includegraphics[width=0.13\textwidth]{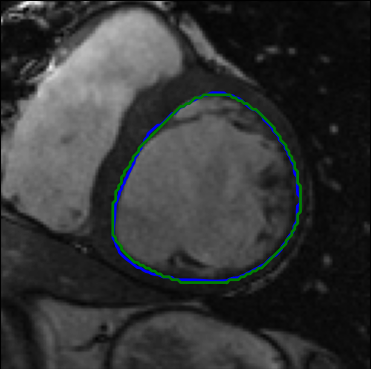}}
    \hspace{3pt}
    
    \stackunder[5pt]{\includegraphics[width=0.13\textwidth]{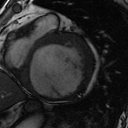}}{ES}
    \hspace{3pt}
    \stackunder[5pt]{\includegraphics[width=0.13\textwidth]{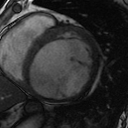}}{ED}
    \hspace{3pt}
    \stackunder[5pt]{\includegraphics[width=0.13\textwidth]{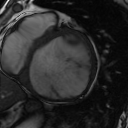}}{Warped ES}
    \hspace{3pt}
    \stackunder[5pt]{ \includegraphics[width=0.13\textwidth]{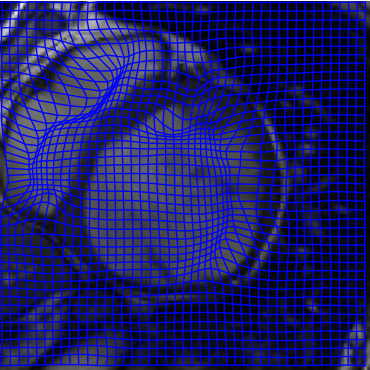}}{DF}
    \hspace{3pt}
    \stackunder[5pt]{\includegraphics[width=0.13\textwidth]{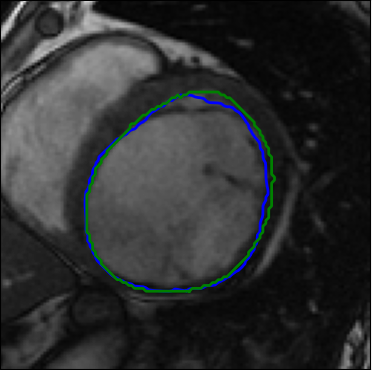}}{GT}

    \begin{minipage}{0.99\linewidth}
		\centering
		\centerline{SCD}\medskip
	\end{minipage}
     \caption{Samples of registered images on the SCD with the corresponding deformation filed grid (DF). End-systolic (ES) frame is the moving image and end-diastolic (ED) frame is the fixed image. The warped ES of each row is shown in the third column. The last column labeled ground truth (GT) displays the true segmentation and the predicted segmentation, which are shown by the green line and blue line respectively.}
     \label{fig6:scd_2d_reg} 
\end{figure*}

\begin{figure*}[htbp]
     \centering
     \subfigure{\includegraphics[width=0.13\textwidth]{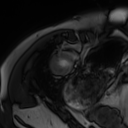}}
    \hspace{3pt}
    \subfigure{\includegraphics[width=0.13\textwidth]{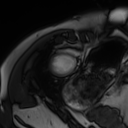}}
      \hspace{3pt}
    \subfigure{\includegraphics[width=0.13\textwidth]{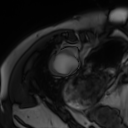}}
      \hspace{3pt}
    \subfigure{\includegraphics[width=0.13\textwidth]{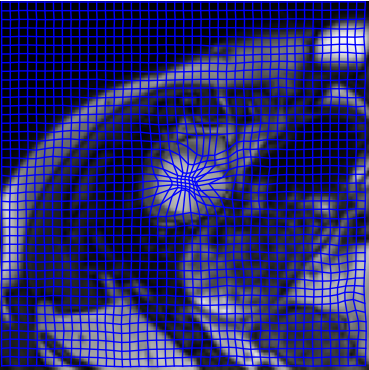}}
      \hspace{3pt}
    \subfigure{\includegraphics[width=0.13\textwidth]{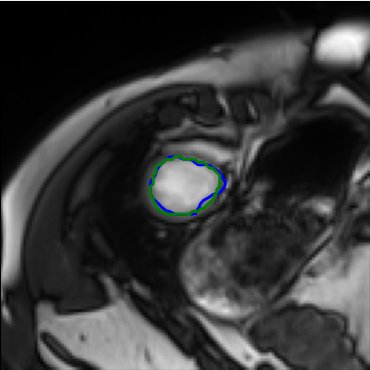}}
    
    \subfigure{\includegraphics[width=0.13\textwidth]{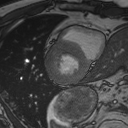}}
    \hspace{3pt}
    \subfigure{\includegraphics[width=0.13\textwidth]{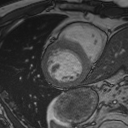}}
      \hspace{3pt}
    \subfigure{\includegraphics[width=0.13\textwidth]{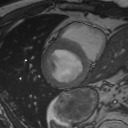}}
      \hspace{3pt}
    \subfigure{\includegraphics[width=0.13\textwidth]{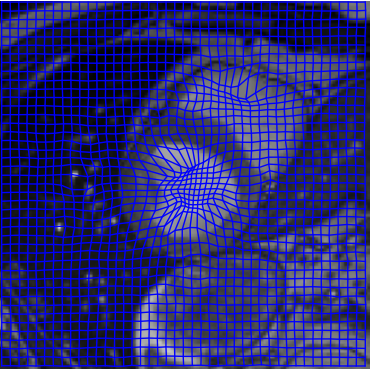}}
      \hspace{3pt}
      \subfigure{\includegraphics[width=0.13\textwidth]{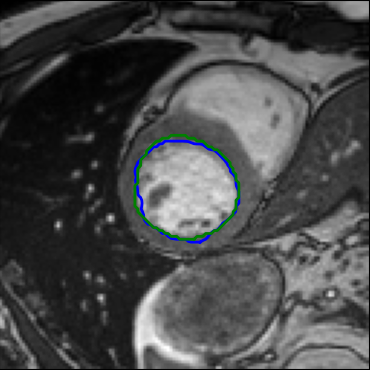}}
     
     \stackunder[5pt]{\includegraphics[width=0.13\textwidth]{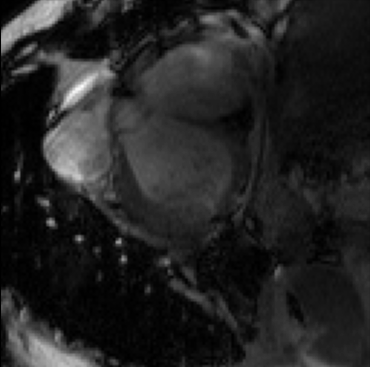}}{ES}
     \hspace{3pt}
     \stackunder[5pt]{\includegraphics[width=0.13\textwidth]{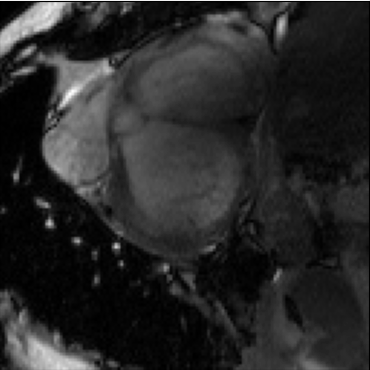}}{ED}
     \hspace{3pt}
     \stackunder[5pt]{\includegraphics[width=0.13\textwidth]{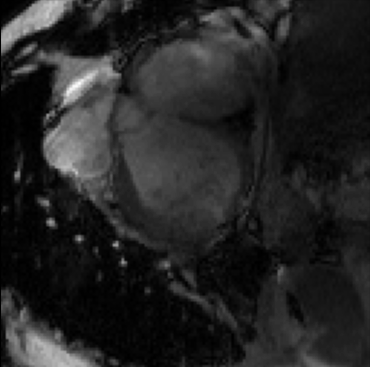}}{Warped ES}
     \hspace{3pt}
     \stackunder[5pt]{\includegraphics[width=0.13\textwidth]{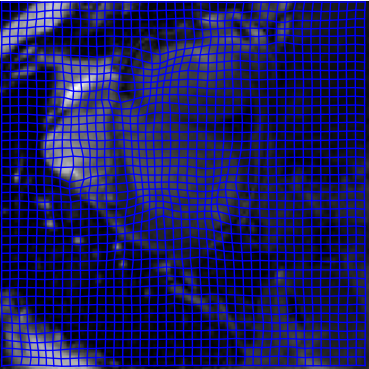}}{DF}
     \hspace{3pt}
     \stackunder[5pt]{\includegraphics[width=0.13\textwidth]{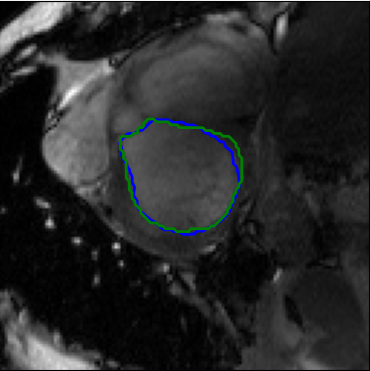}}{GT}
     
     \begin{minipage}{0.99\linewidth}
		\centering
		\centerline{ACDC}\medskip
	\end{minipage}
     \caption{Samples of registered images on the ACDC with the corresponding deformation filed grid (DF). End-systolic (ES) frame is the moving image and end-diastolic (ED) frame is the fixed image. The warped ES of each row is shown in the third column. The last column labeled ground truth (GT) displays the true segmentation and the predicted segmentation, which are shown by the green line and blue line respectively.}
     \label{fig6:acdc_2d_reg}
\end{figure*}
\begin{table*}[htbp]
\centering
\caption{Quantitative evaluation of the results for cardiac MRI registration on the 3D ACDC data set. The following metrics are reported for each method: The Dice score $Dice$ (mean$\pm$ standard deviation), Hausdorff distance $HD$, the percentage of the number of pixels with negative Jacobian determinant $\% |J_\theta|<0$, and reliability $R(0.75)$. Smaller values of $HD$ and larger values of $Dice$ indicate more accurate results. Also the smaller $\% |J_\theta|<0$ indicates less mesh folding. The higher probability values of $R(0.75)$ show that more patients have the dice score higher or equal to $\%0.75$. Values that are highlighted in bold indicate the metric that gave the best performance compared to the other algorithms.}
\label{ch6:ACDCtable_3dreg}
\setlength{\tabcolsep}{3pt}
\begin{tabular}{p{180pt} p{90pt} p{70pt} p{70pt}p{40pt}}
\hline
Method&
Dice& 
HD&
$\% |J_\theta|<0$ &
$R(0.75)$\\
\hline
Undeformed & 0.71 $\pm$ 0.145 & 10.1 & --&-- \\
Demon\cite{yoo2002engineering}& 0.80 $\pm$ 0.17 & 8.3&0.34&0.28 \\
SyN\cite{avants2008symmetric} & 0.80 $\pm$ 0.091 &  8.1&0.17&0.51 \\
LPM\cite{krebs2019learning}& 0.81 $\pm$ 0.085 & 7.3&0.12&0.52\\
LapIRN \cite{mok2020large}& 0.72 $\pm$ 0.162 & 7.4  & 0 & 0.35\\
MM\cite{Krishnaswamy2021} & 0.75 $\pm$ 0.156 & 7.03 & 0 & 0.56 \\
Elastix\cite{marstal2016simpleelastix} & 0.83 $\pm$ 0.161 & 5.75& 0.09& 0.60\\
\setrow{\bfseries}Proposed Method &\setrow{\bfseries}0.84 $\pm$ 0.06 &\setrow{\bfseries} 5.3&\setrow{\bfseries} 0&\setrow{\bfseries} 0.78 \\

\hline
\end{tabular}
\end{table*}
\begin{figure*}[htbp]
     \begin{minipage}{0.9\linewidth}
		\centering
		\centerline{ACDC}\medskip
	\end{minipage}
     \subfigure{\includegraphics[width=0.195\textwidth]{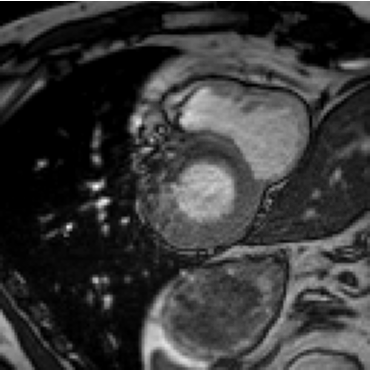}}
     \hspace{2pt}
     \subfigure{\includegraphics[width=0.195\textwidth]{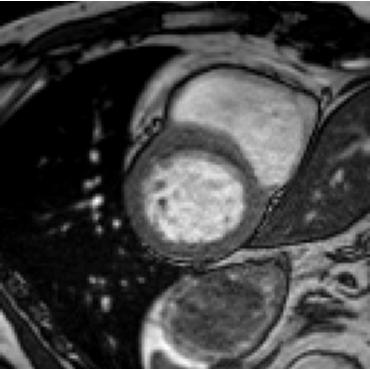}}
     \hspace{2pt}
    \subfigure{\includegraphics[width=0.25\textwidth]{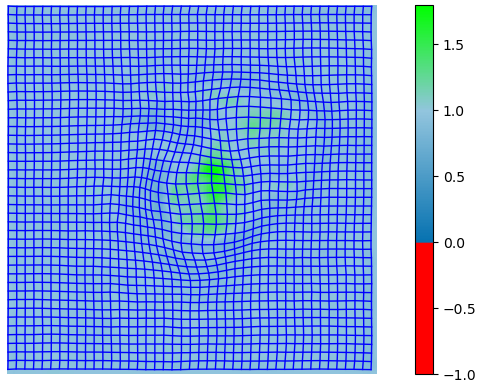}}
     \hspace{2pt}
      \subfigure{\includegraphics[width=0.25\textwidth]{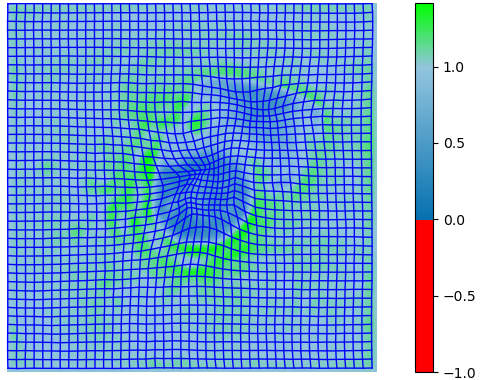}}
     \begin{minipage}{0.9\linewidth}
		\centering
		\centerline{SCD}\medskip
	\end{minipage}
   
     \subfigure{\includegraphics[width=0.195\textwidth]{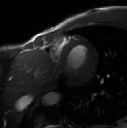}}
     \hspace{2pt}
    \subfigure{\includegraphics[width=0.195\textwidth]{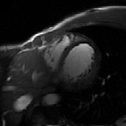}}
     \hspace{2pt}
     \subfigure{\includegraphics[width=0.25\textwidth]{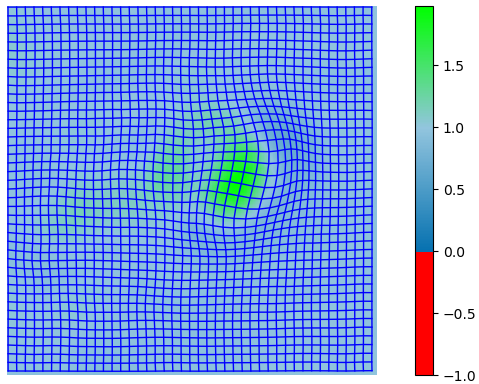}}
     \hspace{2pt}
    \subfigure{\includegraphics[width=0.25\textwidth]{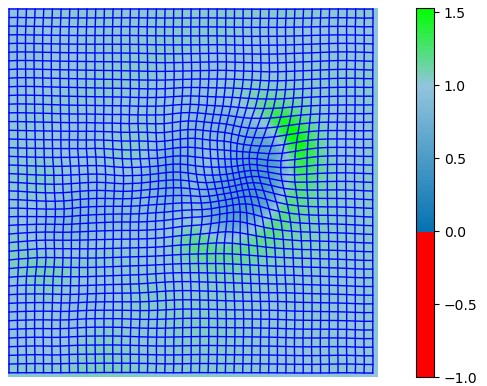}}
      \begin{minipage}{0.9\linewidth}
		\centering
		\centerline{left-2ch}\medskip
	\end{minipage}
     \subfigure{\includegraphics[width=0.195\textwidth]{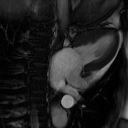}}
     \hspace{2pt}
     \subfigure{\includegraphics[width=0.195\textwidth]{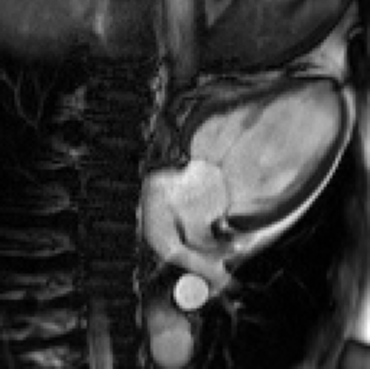}}
     \hspace{2pt}
      \subfigure{\includegraphics[width=0.25\textwidth]{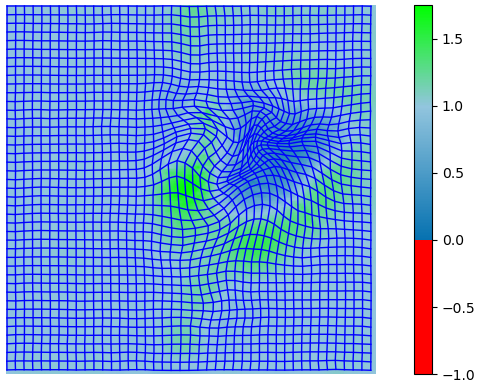}}
     \hspace{2pt}
     \subfigure{\includegraphics[width=0.24\textwidth]{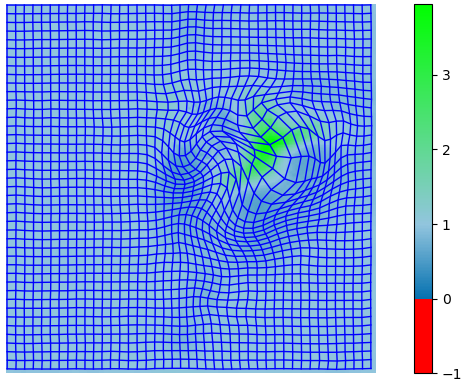}}
      \begin{minipage}{0.9\linewidth}
		\centering
		\centerline{left-3ch}\medskip
	\end{minipage}
        
     \subfigure{\includegraphics[width=0.195\textwidth]{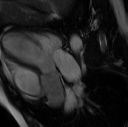}}
     \hspace{2pt}
     \subfigure{\includegraphics[width=0.195\textwidth]{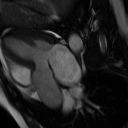}}
     \hspace{2pt}
      \subfigure{\includegraphics[width=0.25\textwidth]{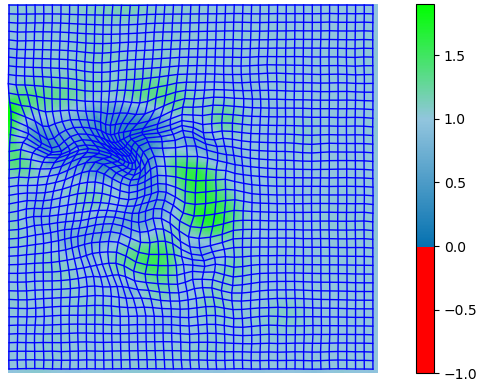}}
     \hspace{2pt}
     \subfigure{\includegraphics[width=0.24\textwidth]{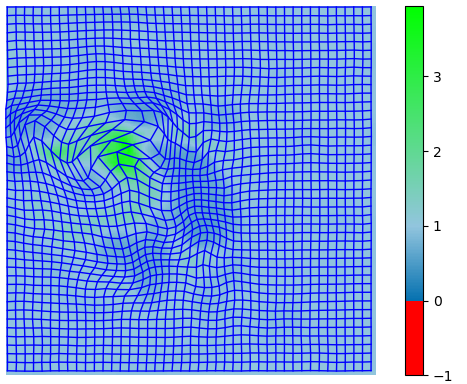}}
      \begin{minipage}{0.9\linewidth}
		\centering
		\centerline{left-4ch}\medskip
	\end{minipage}
        
     \stackunder[5pt]{\includegraphics[width=0.195\textwidth]{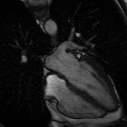}}{ES}
     \hspace{0.55cm}
     \stackunder[5pt]{\includegraphics[width=0.195\textwidth]{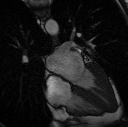}}{ED}
     \hspace{0.6cm}
     \stackunder[5pt]{\includegraphics[width=0.25\textwidth]{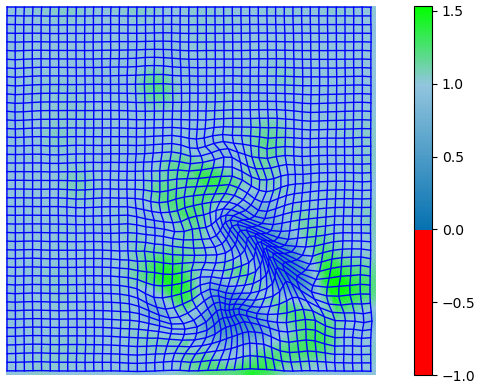}}{Forward DF}
     \hspace{0.53cm}
     \stackunder[5pt]{\includegraphics[width=0.238\textwidth]{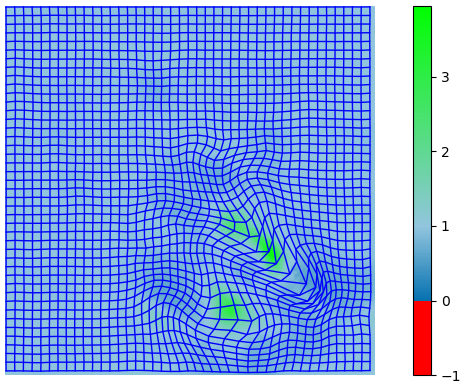}}{Backward DF}
     \hspace{2pt}

	\caption{2D registration results for five example patients, where the first column is the end-systolic image and the second column is the end-diastolic image. The grid deformations in the 3rd column displays the deformation from end-systole to end-diastole, while the last column displays the deformation from end-diastole to end-systole. The color represents the value of the Jacobian determinant, where red indicates values below 0, which is where mesh folding occurs. It can be seen that using the proposed method, no mesh folding occurs.}
	\label{fig6:detJ}
\end{figure*}

\subsection{Baseline Methods}
We compared the performance of the proposed framework with state-of-the-art algorithms, SimpleElastix (Elastix) \cite{marstal2016simpleelastix},(MM) \cite{punithakumar2017gpu, Krishnaswamy2021}, Fast Symmetric Forces Demons (Demons) \cite{mccormick2014itk}, Symmetric Normalization \cite{avants2008symmetric} which are optimization based methods and diffeomorphic learning-based methods LPM \cite{krebs2019learning} and LapIRN \cite{mok2020large}.

\subsubsection{2D Image Registration Results}
Tables \ref{ch6:ACDCtable_2dreg}, \ref{ch6:lefttable_2dreg} and \ref{ch6:Sunnytable_2dreg} provide a summary of the results of the proposed method, the mean and standard deviations of DM, HD, the percentage of the number of pixels with negative Jacobian determinant $\% |J_\theta|<0$, and reliability $R(0.75)$ on the held out test set on ACDC, LA, and SCD data sets, respectively. Figures \ref{fig6:la_2d_reg}, \ref{fig6:scd_2d_reg}, \ref{fig6:acdc_2d_reg} show samples of registered images on the LA, SCD, and ACDC data sets with the corresponding deformation field grid. End-systolic frame is the moving image and end-diastolic frame is the fixed image. The registered image of each row is shown in the third column. Also, the true and predicted segmentation maps are shown by the green and blue line respectively. For each new 2D pair of images, the registration process takes an average of $0.05 \pm 0.03$ seconds on a GPU.

The ACDC data set is originally a 3D data set where a set of 2D axial slices are stacked to form a 3D volume. To evaluate the 2D version of the proposed framework on ACDC, we computed 2D metrics on each slice separately and aggregated the results over all slices to obtain the final values reported in Table \ref{ch6:ACDCtable_2dreg}. 

The presented method shows a better performance among the all
compared methods in all aspects e.g., there is a noticeable difference between the obtain Dice score and Hausdorff distance. As can be seen, the improvement is not just limited to these two parameters, the Jacobian determinant is zero which means there is no folding or twisting in the transformation. This is in contrast to other methods where the determinant Jacobian is non-zero.

		


Figure. \ref{fig6:detJ} shows the end-diastolic and end-systolic images and the determinant of the Jacobian ($|J_\theta|$) with grid overlay for five example patients. As shown in all tables and Figure. \ref{fig6:detJ}, no negative values were observed on the test data for the proposed method which means our approach produced smooth and regular deformations.  

\subsubsection{3D Image Registration results}
The publicly available Automated Cardiac Diagnosis Challenge (ACDC) data set was employed for the evaluation of the proposed 3D-to-3D registration algorithm. Table \ref{ch6:ACDCtable_3dreg} provides a summary of the results of the proposed method on the ACDC data set. The presented method displays a better performance among all the compared methods in all aspects e.g., there is a noticeable difference between the obtained Dice score and Hausdorff distance. Also, the higher probability values of $R(0.75)$ proves that the proposed method is more reliable than the other compared methods since more patients have the dice score higher or equal to $\%0.75$. In addition, similar to 2D version, the Jacobian determinant is also zero in 3D version which means there is no mesh folding in the transformation. The registration process takes an average of $0.07 \pm 0.005$ seconds on a GPU to register an unseen 3D pair of images.

\begin{figure}[htbp]
	\centering
	\includegraphics[width=0.5\textwidth]{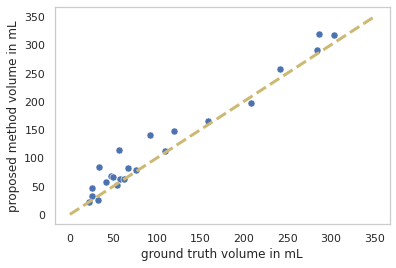}
	\caption{The ground truth volume in mL plotted against the volume from the proposed method, where each patient is represented by a blue dot. The yellow dotted line indicates the y=x line for reference. The Pearson correlation coefficient calculated is 0.98, revealing a high correlation of the proposed method to the ground truth.
}
	\label{fig:volume_corr}
\end{figure}

Figure \ref{fig:volume_corr} displays a correlation plot, where the ground truth volume in mL is plotted against the volume from the proposed method. The clustering of the dots to the reference yellow line indicates the high agreement between the proposed method to the ground truth. The analysis produced a Pearson correlation coefficient of 0.98. 

\subsubsection{Implementation and Parameters Analysis}
The proposed method is implemented in Python programming language using Pytorch module. The network is designed based on a UNet-style architecture \cite{ronneberger2015u} which includes a convolutional layer with 16 filters, three downsampling layers with 32,64,64 convolutional filters and a stride of two, and upsampling convolutional layers with 64,64,32,32,32,16 filters. The Adam optimization with learning rate of $5\times10^{-4}$ is used for all the three datasets. The proposed framework is evaluated on an NVIDIA GeForce GTX 1080 Ti GPU.

To guarantee the diffeomorphism and keep the transformation determinant Jacobian positive, different activation functions are used to apply constraints on $\mu$ and $V(\xi)$ and keep their range in $(0,1)$ and $(-\lambda,+\lambda)$ respectively. Where $\lambda$ can be any value in range of $(1 , \infty)$, we set $\lambda = 10$ in our experiment. Then using the two hyper-parameters, lower bound $\tau_{lb} \in (0,1)$ and upper bound $\tau_{ub} \in (1 , \lambda)$ of the transformation Jacobian determinant $|J_{\theta}|$, the user can control the amount of movement which directly affects the evaluation metrics. By increasing the values of $\tau_{lb}$ and $\tau_{ub}$, each node in a grid (each pixel) can have a larger displacement; however, after a certain point, the results do not change significantly. We vary the precision $\tau_{lb}$, $\tau_{ub}$ and set them to $0.2$, $8.0$ respectively. The chosen values resulted in the best Dice score and HD distance.


\section{Conclusion} 
In this work, we build a principled connection between classical registration methods and recent learning-based approaches. We propose an end-to-end framework for diffeomorphic image registration and derive a learning algorithm that leverages a convolutional neural network and unsupervised learning for fast runtime. To achieve diffeomorphic transforms, we integrate a new parameterization of deformation fields for 2D-to-2D and 3D-to-3D diffeomorphic registration algorithm for the application of MRI cardiac registration, which describe a deformation field with its transformation Jacobian determinant and curl of the end velocity field. It also relaxes the need for an explicit regularization to produce a physically plausible result, as smoothness is implicitly embedded in the solution. Removing explicit regularization makes the need for an empirical trade-off between the similarity term and the regularization term, which may cause bias \cite{beg2005computing}, unnecessary. 

Furthermore, by directly requiring the transformation Jacobian to be positive, the deformation can be ensured to be diffeomorphic. The other desirable constraints also can be enforced within the same framework using an explicit restriction on the transformation Jacobian such as incompressibility constraint. Additionally, the proposed parameterization naturally describes a deformation field in terms of radial and rotational components, making it especially suited for processing cardiac data \cite{noble2002myocardial}. Our algorithm can infer the registration of new image pairs in under a second, which is significantly faster than traditional iterative methods. Compared to recent learning-based methods, our method offers a guarantee of a diffeomorphic transform. 

The proposed algorithm was evaluated on end-diastolic to end-systolic cardiac cine-MRI registration on two publicly available ACDC Challenge \cite{Bernard2018} and Sunnybrook data sets (SCD)\cite{radau2009evaluation} as well as a set of left atrium images obtained from the Mazankowski Alberta Heart Institute. The proposed algorithm is diffeomorphic, allowing it to capture the true deformation of the cardiac tissue. Observing the percentage of voxels with a Jacobian determinant less than zero, most of the other registration methods yielded mesh folding for either the MRI data sets. The presence of mesh folding may result in the inability of these methods to capture the true anatomical motion.

\section*{Acknowledgment}

The authors wish to thank Alberta Innovates for the AICE Concepts funding that supported this research work.
\bibliographystyle{model2-names.bst}\biboptions{authoryear}
\bibliography{refs}

\begin{thebibliography}{47}
\expandafter\ifx\csname natexlab\endcsname\relax\def\natexlab#1{#1}\fi
\providecommand{\url}[1]{\texttt{#1}}
\providecommand{\href}[2]{#2}
\providecommand{\path}[1]{#1}
\providecommand{\DOIprefix}{doi:}
\providecommand{\ArXivprefix}{arXiv:}
\providecommand{\URLprefix}{URL: }
\providecommand{\Pubmedprefix}{pmid:}
\providecommand{\doi}[1]{\href{http://dx.doi.org/#1}{\path{#1}}}
\providecommand{\Pubmed}[1]{\href{pmid:#1}{\path{#1}}}
\providecommand{\bibinfo}[2]{#2}
\ifx\xfnm\relax \def\xfnm[#1]{\unskip,\space#1}\fi
\bibitem[{Ashburner et~al.(1999)Ashburner, Andersson and
  Friston}]{ashburner1999high}
\bibinfo{author}{Ashburner, J.}, \bibinfo{author}{Andersson, J.L.},
  \bibinfo{author}{Friston, K.J.}, \bibinfo{year}{1999}.
\newblock \bibinfo{title}{High-dimensional image registration using symmetric
  priors}.
\newblock \bibinfo{journal}{NeuroImage} \bibinfo{volume}{9},
  \bibinfo{pages}{619--628}.
\bibitem[{Avants et~al.(2008)Avants, Epstein, Grossman and
  Gee}]{avants2008symmetric}
\bibinfo{author}{Avants, B.B.}, \bibinfo{author}{Epstein, C.L.},
  \bibinfo{author}{Grossman, M.}, \bibinfo{author}{Gee, J.C.},
  \bibinfo{year}{2008}.
\newblock \bibinfo{title}{Symmetric diffeomorphic image registration with
  cross-correlation: evaluating automated labeling of elderly and
  neurodegenerative brain}.
\newblock \bibinfo{journal}{Medical image analysis} \bibinfo{volume}{12},
  \bibinfo{pages}{26--41}.
\bibitem[{Balakrishnan et~al.(2018)Balakrishnan, Zhao, Sabuncu, Guttag and
  Dalca}]{balakrishnan2018unsupervised}
\bibinfo{author}{Balakrishnan, G.}, \bibinfo{author}{Zhao, A.},
  \bibinfo{author}{Sabuncu, M.R.}, \bibinfo{author}{Guttag, J.},
  \bibinfo{author}{Dalca, A.V.}, \bibinfo{year}{2018}.
\newblock \bibinfo{title}{An unsupervised learning model for deformable medical
  image registration}, in: \bibinfo{booktitle}{Proceedings of the IEEE
  conference on computer vision and pattern recognition}, pp.
  \bibinfo{pages}{9252--9260}.
\bibitem[{Beg et~al.(2005)Beg, Miller, Trouv{\'e} and
  Younes}]{beg2005computing}
\bibinfo{author}{Beg, M.F.}, \bibinfo{author}{Miller, M.I.},
  \bibinfo{author}{Trouv{\'e}, A.}, \bibinfo{author}{Younes, L.},
  \bibinfo{year}{2005}.
\newblock \bibinfo{title}{Computing large deformation metric mappings via
  geodesic flows of diffeomorphisms}.
\newblock \bibinfo{journal}{International journal of computer vision}
  \bibinfo{volume}{61}, \bibinfo{pages}{139--157}.
\bibitem[{Bernard et~al.(2018a)Bernard, Lalande, Zotti, Cervenansky, Yang,
  Heng, Cetin, Lekadir, Camara, Ballester and Sanroma}]{Bernard2018}
\bibinfo{author}{Bernard, O.}, \bibinfo{author}{Lalande, A.},
  \bibinfo{author}{Zotti, C.}, \bibinfo{author}{Cervenansky, F.},
  \bibinfo{author}{Yang, X.}, \bibinfo{author}{Heng, P.},
  \bibinfo{author}{Cetin, I.}, \bibinfo{author}{Lekadir, K.},
  \bibinfo{author}{Camara, O.}, \bibinfo{author}{Ballester, M.},
  \bibinfo{author}{Sanroma, G.}, \bibinfo{year}{2018}a.
\newblock \bibinfo{title}{Deep learning techniques for automatic {MRI} cardiac
  multi-structures segmentation and diagnosis: Is the problem solved?}
\newblock \bibinfo{journal}{IEEE transactions on medical imaging}
  \bibinfo{volume}{37}, \bibinfo{pages}{2514--25}.
\bibitem[{Bernard et~al.(2018b)Bernard, Lalande, Zotti, Cervenansky, Yang,
  Heng, Cetin, Lekadir, Camara, Ballester et~al.}]{bernard2018deep}
\bibinfo{author}{Bernard, O.}, \bibinfo{author}{Lalande, A.},
  \bibinfo{author}{Zotti, C.}, \bibinfo{author}{Cervenansky, F.},
  \bibinfo{author}{Yang, X.}, \bibinfo{author}{Heng, P.A.},
  \bibinfo{author}{Cetin, I.}, \bibinfo{author}{Lekadir, K.},
  \bibinfo{author}{Camara, O.}, \bibinfo{author}{Ballester, M.A.G.}, et~al.,
  \bibinfo{year}{2018}b.
\newblock \bibinfo{title}{Deep learning techniques for automatic {MRI} cardiac
  multi-structures segmentation and diagnosis: {I}s the problem solved?}
\newblock \bibinfo{journal}{IEEE transactions on medical imaging}
  \bibinfo{volume}{37}, \bibinfo{pages}{2514--2525}.
\bibitem[{Bijnens et~al.(2012)Bijnens, Cikes, Butakoff, Sitges and
  Crispi}]{bijnens2012myocardial}
\bibinfo{author}{Bijnens, B.}, \bibinfo{author}{Cikes, M.},
  \bibinfo{author}{Butakoff, C.}, \bibinfo{author}{Sitges, M.},
  \bibinfo{author}{Crispi, F.}, \bibinfo{year}{2012}.
\newblock \bibinfo{title}{Myocardial motion and deformation: What does it tell
  us and how does it relate to function?}
\newblock \bibinfo{journal}{Fetal diagnosis and therapy} \bibinfo{volume}{32},
  \bibinfo{pages}{5--16}.
\bibitem[{Cao et~al.(2017)Cao, Yang, Zhang, Nie, Kim, Wang and
  Shen}]{cao2017deformable}
\bibinfo{author}{Cao, X.}, \bibinfo{author}{Yang, J.}, \bibinfo{author}{Zhang,
  J.}, \bibinfo{author}{Nie, D.}, \bibinfo{author}{Kim, M.},
  \bibinfo{author}{Wang, Q.}, \bibinfo{author}{Shen, D.}, \bibinfo{year}{2017}.
\newblock \bibinfo{title}{Deformable image registration based on
  similarity-steered {CNN} regression}, in: \bibinfo{booktitle}{International
  Conference on Medical Image Computing and Computer-Assisted Intervention},
  \bibinfo{organization}{Springer}. pp. \bibinfo{pages}{300--308}.
\bibitem[{Chen et~al.(2010)Chen, Goela, Garvin and
  Li}]{chen2010parameterization}
\bibinfo{author}{Chen, H.m.}, \bibinfo{author}{Goela, A.},
  \bibinfo{author}{Garvin, G.J.}, \bibinfo{author}{Li, S.},
  \bibinfo{year}{2010}.
\newblock \bibinfo{title}{A parameterization of deformation fields for
  diffeomorphic image registration and its application to myocardial
  delineation}, in: \bibinfo{booktitle}{International Conference on Medical
  Image Computing and Computer-Assisted Intervention},
  \bibinfo{organization}{Springer}. pp. \bibinfo{pages}{340--348}.
\bibitem[{Cheng(1989)}]{Cheng1989}
\bibinfo{author}{Cheng, D.}, \bibinfo{year}{1989}.
\newblock \bibinfo{title}{Field and wave electromagnetics}.
\newblock \bibinfo{publisher}{Pearson Education India}.
\bibitem[{Dalca et~al.(2018)Dalca, Balakrishnan, Guttag and
  Sabuncu}]{dalca2018unsupervised}
\bibinfo{author}{Dalca, A.V.}, \bibinfo{author}{Balakrishnan, G.},
  \bibinfo{author}{Guttag, J.}, \bibinfo{author}{Sabuncu, M.R.},
  \bibinfo{year}{2018}.
\newblock \bibinfo{title}{Unsupervised learning for fast probabilistic
  diffeomorphic registration}, in: \bibinfo{booktitle}{International Conference
  on Medical Image Computing and Computer-Assisted Intervention},
  \bibinfo{organization}{Springer}. pp. \bibinfo{pages}{729--738}.
\bibitem[{Dice(1945)}]{dice1945measures}
\bibinfo{author}{Dice, L.R.}, \bibinfo{year}{1945}.
\newblock \bibinfo{title}{Measures of the amount of ecologic association
  between species}.
\newblock \bibinfo{journal}{Ecology} \bibinfo{volume}{26},
  \bibinfo{pages}{297--302}.
\bibitem[{Garreau et~al.(2006)Garreau, Simon, Boulmier, Coatrieux and
  Le~Breton}]{garreau2006assessment}
\bibinfo{author}{Garreau, M.}, \bibinfo{author}{Simon, A.},
  \bibinfo{author}{Boulmier, D.}, \bibinfo{author}{Coatrieux, J.L.},
  \bibinfo{author}{Le~Breton, H.}, \bibinfo{year}{2006}.
\newblock \bibinfo{title}{Assessment of left ventricular function in cardiac
  msct imaging by a 4d hierarchical surface-volume matching process}.
\newblock \bibinfo{journal}{International Journal of Biomedical Imaging}
  \bibinfo{volume}{2006}.
\bibitem[{Haber and Modersitzki(2004)}]{haber2004numerical}
\bibinfo{author}{Haber, E.}, \bibinfo{author}{Modersitzki, J.},
  \bibinfo{year}{2004}.
\newblock \bibinfo{title}{Numerical methods for volume preserving image
  registration}.
\newblock \bibinfo{journal}{Inverse problems} \bibinfo{volume}{20},
  \bibinfo{pages}{1621}.
\bibitem[{Han et~al.(2021)Han, Uneri, Vijayan, Wu, Vagdargi, Sheth, Vogt,
  Kleinszig, Osgood and Siewerdsen}]{han2021fracture}
\bibinfo{author}{Han, R.}, \bibinfo{author}{Uneri, A.},
  \bibinfo{author}{Vijayan, R.C.}, \bibinfo{author}{Wu, P.},
  \bibinfo{author}{Vagdargi, P.}, \bibinfo{author}{Sheth, N.},
  \bibinfo{author}{Vogt, S.}, \bibinfo{author}{Kleinszig, G.},
  \bibinfo{author}{Osgood, G.}, \bibinfo{author}{Siewerdsen, J.H.},
  \bibinfo{year}{2021}.
\newblock \bibinfo{title}{Fracture reduction planning and guidance in
  orthopaedic trauma surgery via multi-body image registration}.
\newblock \bibinfo{journal}{Medical image analysis} \bibinfo{volume}{68},
  \bibinfo{pages}{101917}.
\bibitem[{Haskins et~al.(2019)Haskins, Kruger and Yan}]{haskins2019deep}
\bibinfo{author}{Haskins, G.}, \bibinfo{author}{Kruger, U.},
  \bibinfo{author}{Yan, P.}, \bibinfo{year}{2019}.
\newblock \bibinfo{title}{Deep learning in medical image registration: A
  survey}.
\newblock \bibinfo{journal}{arXiv preprint arXiv:1903.02026} .
\bibitem[{Huttenlocher et~al.(1993)Huttenlocher, Klanderman and
  Rucklidge}]{huttenlocher1993comparing}
\bibinfo{author}{Huttenlocher, D.P.}, \bibinfo{author}{Klanderman, G.A.},
  \bibinfo{author}{Rucklidge, W.J.}, \bibinfo{year}{1993}.
\newblock \bibinfo{title}{Comparing images using the \textbf{hausdorff}
  distance}.
\newblock \bibinfo{journal}{IEEE Transactions on pattern analysis and machine
  intelligence} \bibinfo{volume}{15}, \bibinfo{pages}{850--863}.
\bibitem[{Jaderberg et~al.(2015)Jaderberg, Simonyan, Zisserman
  et~al.}]{jaderberg2015spatial}
\bibinfo{author}{Jaderberg, M.}, \bibinfo{author}{Simonyan, K.},
  \bibinfo{author}{Zisserman, A.}, et~al., \bibinfo{year}{2015}.
\newblock \bibinfo{title}{Spatial transformer networks}, in:
  \bibinfo{booktitle}{Advances in neural information processing systems}, pp.
  \bibinfo{pages}{2017--2025}.
\bibitem[{Krebs et~al.(2019)Krebs, e~Delingette, Mailh{\'e}, Ayache and
  Mansi}]{krebs2019learning}
\bibinfo{author}{Krebs, J.}, \bibinfo{author}{e~Delingette, H.},
  \bibinfo{author}{Mailh{\'e}, B.}, \bibinfo{author}{Ayache, N.},
  \bibinfo{author}{Mansi, T.}, \bibinfo{year}{2019}.
\newblock \bibinfo{title}{Learning a probabilistic model for diffeomorphic
  registration}.
\newblock \bibinfo{journal}{IEEE transactions on medical imaging} .
\bibitem[{Krebs et~al.(2018)Krebs, Mansi, Mailh{\'e}, Ayache and
  Delingette}]{krebs2018unsupervised}
\bibinfo{author}{Krebs, J.}, \bibinfo{author}{Mansi, T.},
  \bibinfo{author}{Mailh{\'e}, B.}, \bibinfo{author}{Ayache, N.},
  \bibinfo{author}{Delingette, H.}, \bibinfo{year}{2018}.
\newblock \bibinfo{title}{Unsupervised probabilistic deformation modeling for
  robust diffeomorphic registration}, in: \bibinfo{booktitle}{Deep Learning in
  Medical Image Analysis and Multimodal Learning for Clinical Decision
  Support}. \bibinfo{publisher}{Springer}, pp. \bibinfo{pages}{101--109}.
\bibitem[{Krishnaswamy(2021)}]{Krishnaswamy2021}
\bibinfo{author}{Krishnaswamy, D.}, \bibinfo{year}{2021}.
\newblock \bibinfo{title}{A Diffeomorphic 3D-to-3D Registration Algorithm for
  the Segmentation of the Left Ventricle in Ultrasound Sequences}.
\newblock Ph.D. thesis. University of Alberta.
\bibitem[{Liu(2006)}]{liu2006new}
\bibinfo{author}{Liu, J.}, \bibinfo{year}{2006}.
\newblock \bibinfo{title}{New development of the deformation method}.
\newblock \bibinfo{publisher}{The University of Texas at Arlington}.
\bibitem[{Liu et~al.(2021)Liu, Aviles-Rivero, Ji and
  Sch{\"o}nlieb}]{liu2021rethinking}
\bibinfo{author}{Liu, J.}, \bibinfo{author}{Aviles-Rivero, A.I.},
  \bibinfo{author}{Ji, H.}, \bibinfo{author}{Sch{\"o}nlieb, C.B.},
  \bibinfo{year}{2021}.
\newblock \bibinfo{title}{Rethinking medical image reconstruction via shape
  prior, going deeper and faster: Deep joint indirect registration and
  reconstruction}.
\newblock \bibinfo{journal}{Medical Image Analysis} \bibinfo{volume}{68},
  \bibinfo{pages}{101930}.
\bibitem[{Mansi et~al.(2011)Mansi, Pennec, Sermesant, Delingette and
  Ayache}]{mansi2011ilogdemons}
\bibinfo{author}{Mansi, T.}, \bibinfo{author}{Pennec, X.},
  \bibinfo{author}{Sermesant, M.}, \bibinfo{author}{Delingette, H.},
  \bibinfo{author}{Ayache, N.}, \bibinfo{year}{2011}.
\newblock \bibinfo{title}{{iLogDemons: A} demons-based registration algorithm
  for tracking incompressible elastic biological tissues}.
\newblock \bibinfo{journal}{International journal of computer vision}
  \bibinfo{volume}{92}, \bibinfo{pages}{92--111}.
\bibitem[{Marstal et~al.(2016)Marstal, Berendsen, Staring and
  Klein}]{marstal2016simpleelastix}
\bibinfo{author}{Marstal, K.}, \bibinfo{author}{Berendsen, F.},
  \bibinfo{author}{Staring, M.}, \bibinfo{author}{Klein, S.},
  \bibinfo{year}{2016}.
\newblock \bibinfo{title}{Simpleelastix: A user-friendly, multi-lingual library
  for medical image registration}, in: \bibinfo{booktitle}{Proceedings of the
  IEEE conference on computer vision and pattern recognition workshops}, pp.
  \bibinfo{pages}{134--142}.
\bibitem[{McCormick et~al.(2014)McCormick, Liu, Ibanez, Jomier and
  Marion}]{mccormick2014itk}
\bibinfo{author}{McCormick, M.M.}, \bibinfo{author}{Liu, X.},
  \bibinfo{author}{Ibanez, L.}, \bibinfo{author}{Jomier, J.},
  \bibinfo{author}{Marion, C.}, \bibinfo{year}{2014}.
\newblock \bibinfo{title}{{ITK:} enabling reproducible research and open
  science}.
\newblock \bibinfo{journal}{Frontiers in neuroinformatics} \bibinfo{volume}{8},
  \bibinfo{pages}{13}.
\bibitem[{Mok and Chung(2020)}]{mok2020large}
\bibinfo{author}{Mok, T.C.}, \bibinfo{author}{Chung, A.}, \bibinfo{year}{2020}.
\newblock \bibinfo{title}{Large deformation diffeomorphic image registration
  with laplacian pyramid networks}, in: \bibinfo{booktitle}{International
  Conference on Medical Image Computing and Computer-Assisted Intervention},
  \bibinfo{organization}{Springer}. pp. \bibinfo{pages}{211--221}.
\bibitem[{Noble et~al.(2002)Noble, Hill, Breeuwer, Schnabel, Hawkes, Gerritsen
  and Razavi}]{noble2002myocardial}
\bibinfo{author}{Noble, N.M.}, \bibinfo{author}{Hill, D.L.},
  \bibinfo{author}{Breeuwer, M.}, \bibinfo{author}{Schnabel, J.A.},
  \bibinfo{author}{Hawkes, D.J.}, \bibinfo{author}{Gerritsen, F.A.},
  \bibinfo{author}{Razavi, R.}, \bibinfo{year}{2002}.
\newblock \bibinfo{title}{Myocardial delineation via registration in a polar
  coordinate system}, in: \bibinfo{booktitle}{International Conference on
  Medical Image Computing and Computer-Assisted Intervention},
  \bibinfo{organization}{Springer}. pp. \bibinfo{pages}{651--658}.
\bibitem[{Punithakumar et~al.(2013)Punithakumar, Ayed, Islam, Goela, Ross,
  Chong and Li}]{punithakumar2013regional}
\bibinfo{author}{Punithakumar, K.}, \bibinfo{author}{Ayed, I.B.},
  \bibinfo{author}{Islam, A.}, \bibinfo{author}{Goela, A.},
  \bibinfo{author}{Ross, I.G.}, \bibinfo{author}{Chong, J.},
  \bibinfo{author}{Li, S.}, \bibinfo{year}{2013}.
\newblock \bibinfo{title}{Regional heart motion abnormality detection: An
  information theoretic approach}.
\newblock \bibinfo{journal}{Medical image analysis} \bibinfo{volume}{17},
  \bibinfo{pages}{311--324}.
\bibitem[{Punithakumar et~al.(2017)Punithakumar, Boulanger and
  Noga}]{punithakumar2017gpu}
\bibinfo{author}{Punithakumar, K.}, \bibinfo{author}{Boulanger, P.},
  \bibinfo{author}{Noga, M.}, \bibinfo{year}{2017}.
\newblock \bibinfo{title}{A {GPU}-accelerated deformable image registration
  algorithm with applications to right ventricular segmentation}.
\newblock \bibinfo{journal}{IEEE Access} \bibinfo{volume}{5},
  \bibinfo{pages}{20374--20382}.
\bibitem[{Punithakumar et~al.(2015)Punithakumar, Noga, Ayed and
  Boulanger}]{punithakumar2015right}
\bibinfo{author}{Punithakumar, K.}, \bibinfo{author}{Noga, M.},
  \bibinfo{author}{Ayed, I.B.}, \bibinfo{author}{Boulanger, P.},
  \bibinfo{year}{2015}.
\newblock \bibinfo{title}{Right ventricular segmentation in cardiac mri with
  moving mesh correspondences}.
\newblock \bibinfo{journal}{Computerized Medical Imaging and Graphics}
  \bibinfo{volume}{43}, \bibinfo{pages}{15--25}.
\bibitem[{Radau et~al.(2009)Radau, Lu, Connelly, Paul, Dick and
  Wright}]{radau2009evaluation}
\bibinfo{author}{Radau, P.}, \bibinfo{author}{Lu, Y.},
  \bibinfo{author}{Connelly, K.}, \bibinfo{author}{Paul, G.},
  \bibinfo{author}{Dick, A.}, \bibinfo{author}{Wright, G.},
  \bibinfo{year}{2009}.
\newblock \bibinfo{title}{Evaluation framework for algorithms segmenting short
  axis cardiac {MRI}}.
\newblock \bibinfo{journal}{The MIDAS Journal-Cardiac MR Left Ventricle
  Segmentation Challenge} \bibinfo{volume}{49}.
\bibitem[{Roh{\'e} et~al.(2017)Roh{\'e}, Datar, Heimann, Sermesant and
  Pennec}]{rohe2017svf}
\bibinfo{author}{Roh{\'e}, M.M.}, \bibinfo{author}{Datar, M.},
  \bibinfo{author}{Heimann, T.}, \bibinfo{author}{Sermesant, M.},
  \bibinfo{author}{Pennec, X.}, \bibinfo{year}{2017}.
\newblock \bibinfo{title}{Svf-net: learning deformable image registration using
  shape matching}, in: \bibinfo{booktitle}{International conference on medical
  image computing and computer-assisted intervention},
  \bibinfo{organization}{Springer}. pp. \bibinfo{pages}{266--274}.
\bibitem[{Rohlfing(2011)}]{rohlfing2011image}
\bibinfo{author}{Rohlfing, T.}, \bibinfo{year}{2011}.
\newblock \bibinfo{title}{Image similarity and tissue overlaps as surrogates
  for image registration accuracy: widely used but unreliable}.
\newblock \bibinfo{journal}{IEEE transactions on medical imaging}
  \bibinfo{volume}{31}, \bibinfo{pages}{153--163}.
\bibitem[{Ronneberger et~al.(2015)Ronneberger, Fischer and
  Brox}]{ronneberger2015u}
\bibinfo{author}{Ronneberger, O.}, \bibinfo{author}{Fischer, P.},
  \bibinfo{author}{Brox, T.}, \bibinfo{year}{2015}.
\newblock \bibinfo{title}{U-net: Convolutional networks for biomedical image
  segmentation}, in: \bibinfo{booktitle}{International Conference on Medical
  image computing and computer-assisted intervention},
  \bibinfo{organization}{Springer}. pp. \bibinfo{pages}{234--241}.
\bibitem[{Sang et~al.(2020)Sang, Xing, Wu and Ruan}]{sang2020imposing}
\bibinfo{author}{Sang, Y.}, \bibinfo{author}{Xing, X.}, \bibinfo{author}{Wu,
  Y.}, \bibinfo{author}{Ruan, D.}, \bibinfo{year}{2020}.
\newblock \bibinfo{title}{Imposing implicit feasibility constraints on
  deformable image registration using a statistical generative model}, in:
  \bibinfo{booktitle}{Medical Imaging 2020: Image Processing},
  \bibinfo{organization}{International Society for Optics and Photonics}. p.
  \bibinfo{pages}{113132V}.
\bibitem[{Sheikhjafari et~al.(2018)Sheikhjafari, Noga, Punithakumar and
  Ray}]{sheikhjafari2018unsupervised}
\bibinfo{author}{Sheikhjafari, A.}, \bibinfo{author}{Noga, M.},
  \bibinfo{author}{Punithakumar, K.}, \bibinfo{author}{Ray, N.},
  \bibinfo{year}{2018}.
\newblock \bibinfo{title}{Unsupervised deformable image registration with fully
  connected generative neural network}, in: \bibinfo{booktitle}{Medical Imaging
  with Deep Learning}.
\bibitem[{Sheikhjafari et~al.(2022)Sheikhjafari, Noga, Punithakumar and
  Ray}]{sheikhjafari2022training}
\bibinfo{author}{Sheikhjafari, A.}, \bibinfo{author}{Noga, M.},
  \bibinfo{author}{Punithakumar, K.}, \bibinfo{author}{Ray, N.},
  \bibinfo{year}{2022}.
\newblock \bibinfo{title}{A training-free recursive multiresolution framework
  for diffeomorphic deformable image registration}.
\newblock \bibinfo{journal}{Applied Intelligence} , \bibinfo{pages}{1--10}.
\bibitem[{Sheikhjafari et~al.(2015a)Sheikhjafari, Talebi and
  Zareinejad}]{sheikhjafari20153d}
\bibinfo{author}{Sheikhjafari, A.}, \bibinfo{author}{Talebi, H.},
  \bibinfo{author}{Zareinejad, M.}, \bibinfo{year}{2015}a.
\newblock \bibinfo{title}{3d visual stabilization for robotic-assisted beating
  heart surgery using a thin-plate spline deformable model}, in:
  \bibinfo{booktitle}{2015 3rd RSI International Conference on Robotics and
  Mechatronics (ICROM)}, \bibinfo{organization}{IEEE}. pp.
  \bibinfo{pages}{743--748}.
\bibitem[{Sheikhjafari et~al.(2015b)Sheikhjafari, Talebi and
  Zareinejad}]{sheikhjafari2015robust}
\bibinfo{author}{Sheikhjafari, A.}, \bibinfo{author}{Talebi, H.A.},
  \bibinfo{author}{Zareinejad, M.}, \bibinfo{year}{2015}b.
\newblock \bibinfo{title}{Robust and efficient 3d motion tracking in robotic
  assisted beating heart surgery}, in: \bibinfo{booktitle}{2015 IEEE
  International Conference on Robotics and Biomimetics (ROBIO)},
  \bibinfo{organization}{IEEE}. pp. \bibinfo{pages}{1828--1833}.
\bibitem[{Vercauteren et~al.(2008)Vercauteren, Pennec, Perchant and
  Ayache}]{vercauteren2008symmetric}
\bibinfo{author}{Vercauteren, T.}, \bibinfo{author}{Pennec, X.},
  \bibinfo{author}{Perchant, A.}, \bibinfo{author}{Ayache, N.},
  \bibinfo{year}{2008}.
\newblock \bibinfo{title}{Symmetric log-domain diffeomorphic registration: A
  demons-based approach}, in: \bibinfo{booktitle}{International conference on
  medical image computing and computer-assisted intervention},
  \bibinfo{organization}{Springer}. pp. \bibinfo{pages}{754--761}.
\bibitem[{de~Vos et~al.(2019)de~Vos, Berendsen, Viergever, Sokooti, Staring and
  I{\v{s}}gum}]{de2019deep}
\bibinfo{author}{de~Vos, B.D.}, \bibinfo{author}{Berendsen, F.F.},
  \bibinfo{author}{Viergever, M.A.}, \bibinfo{author}{Sokooti, H.},
  \bibinfo{author}{Staring, M.}, \bibinfo{author}{I{\v{s}}gum, I.},
  \bibinfo{year}{2019}.
\newblock \bibinfo{title}{A deep learning framework for unsupervised affine and
  deformable image registration}.
\newblock \bibinfo{journal}{Medical image analysis} \bibinfo{volume}{52},
  \bibinfo{pages}{128--143}.
\bibitem[{de~Vos et~al.(2017)de~Vos, Berendsen, Viergever, Staring and
  I{\v{s}}gum}]{de2017end}
\bibinfo{author}{de~Vos, B.D.}, \bibinfo{author}{Berendsen, F.F.},
  \bibinfo{author}{Viergever, M.A.}, \bibinfo{author}{Staring, M.},
  \bibinfo{author}{I{\v{s}}gum, I.}, \bibinfo{year}{2017}.
\newblock \bibinfo{title}{End-to-end unsupervised deformable image registration
  with a convolutional neural network}, in: \bibinfo{booktitle}{Deep Learning
  in Medical Image Analysis and Multimodal Learning for Clinical Decision
  Support}. \bibinfo{publisher}{Springer}, pp. \bibinfo{pages}{204--212}.
\bibitem[{Wu et~al.(2018)Wu, Bailey, Rasoulinejad and Li}]{wu2018automated}
\bibinfo{author}{Wu, H.}, \bibinfo{author}{Bailey, C.},
  \bibinfo{author}{Rasoulinejad, P.}, \bibinfo{author}{Li, S.},
  \bibinfo{year}{2018}.
\newblock \bibinfo{title}{Automated comprehensive adolescent idiopathic
  scoliosis assessment using mvc-net}.
\newblock \bibinfo{journal}{Medical image analysis} \bibinfo{volume}{48},
  \bibinfo{pages}{1--11}.
\bibitem[{Yoo et~al.(2002)Yoo, Ackerman, Lorensen, Schroeder, Chalana, Aylward,
  Metaxas and Whitaker}]{yoo2002engineering}
\bibinfo{author}{Yoo, T.S.}, \bibinfo{author}{Ackerman, M.J.},
  \bibinfo{author}{Lorensen, W.E.}, \bibinfo{author}{Schroeder, W.},
  \bibinfo{author}{Chalana, V.}, \bibinfo{author}{Aylward, S.},
  \bibinfo{author}{Metaxas, D.}, \bibinfo{author}{Whitaker, R.},
  \bibinfo{year}{2002}.
\newblock \bibinfo{title}{Engineering and algorithm design for an image
  processing api: a technical report on itk-the insight toolkit}, in:
  \bibinfo{booktitle}{Medicine Meets Virtual Reality 02/10}.
  \bibinfo{publisher}{IOS press}, pp. \bibinfo{pages}{586--592}.
\bibitem[{Zhang and Fletcher(2015)}]{zhang2015finite}
\bibinfo{author}{Zhang, M.}, \bibinfo{author}{Fletcher, P.T.},
  \bibinfo{year}{2015}.
\newblock \bibinfo{title}{Finite-dimensional lie algebras for fast
  diffeomorphic image registration}, in: \bibinfo{booktitle}{International
  Conference on Information Processing in Medical Imaging},
  \bibinfo{organization}{Springer}. pp. \bibinfo{pages}{249--260}.
\bibitem[{Zhou(2006)}]{zhou2006uniqueness}
\bibinfo{author}{Zhou, X.}, \bibinfo{year}{2006}.
\newblock \bibinfo{title}{On uniqueness theorem of a vector function}.
\newblock \bibinfo{journal}{Progress in Electromagnetics Research}
  \bibinfo{volume}{65}, \bibinfo{pages}{93--102}.

\end{thebibliography}


\end{document}